\documentclass[a4paper,11pt]{article}

\pdfoutput=1
\usepackage{jcappub}
\usepackage{graphicx}
\usepackage{array}
\usepackage{tabularx}
\usepackage{float}
\usepackage{xcolor}
\usepackage{physics}
\usepackage{slashed}
\usepackage[font=small, labelfont=bf]{caption}
\usepackage[labelformat=simple]{subcaption}

\usepackage{amsmath, amsthm, amssymb, amsfonts}
\usepackage{multirow}

\definecolor{myblue}{rgb}{0.3, 0.5, 0.7}
\definecolor{myyellow}{rgb}{0.85, 0.72, 0.37}

\def\beq{\begin{align}}
\def\eeq{\end{align}}
\newcommand{\bi}{\begin{itemize}}
\newcommand{\ei}{\end{itemize}}
\newcommand{\ben}{\begin{enumerate}}
\newcommand{\een}{\end{enumerate}}
\newcommand{\be}{\begin{equation}}
\newcommand{\ee}{\end{equation}}
\newcommand{\bea}{\begin{eqnarray}}
\newcommand{\eea}{\end{eqnarray}}

\newcommand{\Kahler}{K\"ahler~}
\newcommand{\V}{\mathcal{V}}
\newcommand{\K}{\mathcal{K}}

\renewcommand{\O}{\mathcal{O}}

\def\GW{{\scriptscriptstyle GW}}

\begin{document}

\title{Dynamics of Cosmic Superstrings and the Overshoot Problem}

\author[a,b]{Luca Brunelli,}
\author[a,b]{Michele Cicoli,} 
\author[a,b]{Francisco G. Pedro}
\affiliation[a]{Dipartimento di Fisica e Astronomia, Universit\`a di Bologna, via Irnerio 46, 40126 Bologna, Italy}
\affiliation[b]{INFN, Sezione di Bologna, viale Berti Pichat 6/2, 40127 Bologna, Italy}

\emailAdd{l.brunelli@unibo.it}
\emailAdd{michele.cicoli@unibo.it}
\emailAdd{francisco.soares@unibo.it}

\abstract{  
We exploit the techniques of dynamical systems to study the cosmological evolution of cosmic fundamental strings and effective strings arising from branes wrapped on internal cycles. We also include the whole potential of the volume modulus characterised by an early time run-away towards a late time minimum. We analyse the overshoot problem with and without radiation, and find that the presence of an initial population of strings arising from NS5-branes wrapped around 4-cycles is enough to ensure that the modulus stabilises in its late time minimum, even in the absence of radiation. The reason is the transfer of energy between the modulus and the effective strings caused by the fact that their tension depends on the volume modulus. Interestingly, we find that the energy density of cosmic superstrings is generically very large when the modulus is oscillating around its minimum, opening up the possibility of a detectable gravitational wave signal. We also find no evidence of an efficient resonant enhancement of cosmic superstrings due to an oscillating tension in the late time minimum.}

\maketitle 

\tableofcontents

\section{Introduction}

Early universe cosmology is a very promising arena to probe UV complete theories like string theory. This involves the study of trans-Planckian field excursions during epochs of large field inflation and kination, or the gravitational decay of long-lived moduli (see \cite{Cicoli:2023opf} for a recent review of different aspects of string cosmology). 

In this paper we focus on two features that typically arise at the end of string inflation (see for example brane-antibrane inflation \cite{Burgess:2001fx, Dvali:2001fw, Kachru:2003sx, Baumann:2007ah, Baumann:2007np,Cicoli:2024bwq} or volume modulus inflation \cite{Conlon:2008cj,Cicoli:2015wja}): a volume modulus that rolls over several Planck units towards its late time minimum, and a population of cosmic superstrings with a modulus-dependent tension that varies with time.

A notorious issue of string models of the early universe is the \emph{overshoot problem} \cite{Brustein:1992nk}, that is the tendency of the volume mode to run over the barrier that separates its minimum from the decompactification limit, if the modulus is released with a very high energy after the end of inflation. A known solution relies on the presence of a background fluid, like radiation or matter, which redshifts slower than the modulus kinetic energy and yields additional Hubble friction to the modulus dynamics \cite{Kaloper:1991mq, Barreiro:1998aj, Brustein:2004jp, Kaloper:2004yj, Battefeld:2005av,Itzhaki:2007nk, Conlon:2008cj, Acharya:2008bk}. However, especially in models where the volume mode is rolling before reheating, motivating the microscopic origin of enough radiation to prevent overshooting can be non-trivial \cite{Conlon:2022pnx,Mosny:2025cyd}.

Recently, the powerful techniques of dynamical systems have been exploited to explore in detail the cosmological evolution of systems involving rolling scalars, cosmic superstrings and additional background fluids. Previous studies discovered that, during kination, the physical size of loops of fundamental strings (hereafter referred to as F-strings) \cite{Conlon:2024uob} and effective strings arising from D3- and NS5-branes wrapped around fibration cycles \cite{Brunelli:2025ems} can grow faster than the scale factor and form a cosmic superstring network. Interestingly, this growth can occur for F-strings also for scaling fixed points in the presence of a background fluid with an appropriate equation of state parameter \cite{Brunelli:2025ems}. Moreover, ref. \cite{SanchezGonzalez:2025uco} took into account the energy density in cosmic string loops and revealed, among other things, the existence of a new attractor fixed point (a loop tracker) where the energy density of the universe is dominated by the cosmic superstrings. 

In this paper we extend the analysis of these cosmological systems considering the whole potential of the rolling volume modulus of type IIB LVS scenarios \cite{Balasubramanian:2005zx, Cicoli:2008va}, which includes both an early-time runaway and a late-time minimum with a barrier against decompactification. We find that, without enough radiation, a model with F-strings would face an overshoot problem. On the other hand, as soon as an initial population of effective strings arising from NS5-branes wrapped around 4-cycles is present, the overshoot problem is avoided, even in the absence of radiation. Moreover, the cosmic superstrings constitute around $97\%$ of the total energy density in the loop tracker fixed point that precedes the final evolution of the field towards the minimum of its potential where the energy density of cosmic superstrings can be as high as about $50\%$. Such large values of energy densities imply the potential generation of a detectable gravitational wave (GW) signal at high frequencies from the decay of cosmic superstrings in the early universe. The GW spectrum is expected to be peaked in the GHz regime and be diluted by the subsequent epoch of matter domination caused by the oscillations of the modulus around the minimum of its potential \cite{Conlon:2025mqt}.\footnote{In string cosmology, there are several mechanisms for producing GWs, such as primordial tensor modes during inflation or secondary GWs sourced by large curvature or isocurvature perturbations; however, these typically lie at much lower frequencies.}

\section{Strings with time-dependent tension}
\label{sec:time dependence}

We  start by giving a brief review of the derivation of the equations of motion for  a string loop with time-dependent tension following \cite{Emond:2021vts,Conlon:2024uob,Revello:2024gwa}.
The Nambu-Goto (NG) action for a string with a tension $\mu$ dependent on the spacetime coordinates $x^\mu$ is given by:
\begin{equation}
\label{eq:NG}
S_{\rm NG}=-\int d^2 \sigma \, \mu(x) \,  \sqrt{-\gamma}\,. 
\end{equation}
Here $\sigma^a$ are the worldsheet coordinates and $\gamma = \det (\gamma_{ab})$ where $\gamma_{ab}$ is the pullback of the spacetime metric onto the worldsheet: 
\begin{equation}
\gamma_{ab}= \frac{\partial x^\mu}{\partial \sigma^a} \frac{\partial x^\nu}{\partial \sigma^b}\, g_{\mu\nu}\,.
\end{equation}
Varying \eqref{eq:NG} with respect to $x^\mu$ yields the following equations of motion \cite{Emond:2021vts}:
\begin{equation}
    \frac{1}{\mu \sqrt{-\gamma}}\, \partial_a(\mu \sqrt{-\gamma}\gamma^{ab}x^\mu_{,b})+\Gamma^\mu_{\nu\rho} \gamma^{ab}x^\nu_{,a}x^\rho_{,b}-\frac{\partial^\mu \mu}{\mu}=0\,.
\end{equation}
We can then gauge-fix the worldsheet coordinates so that $x^0=\sigma^0$ and $\frac{\partial \vec{x}}{\partial \sigma^1}\cdot\frac{\partial\vec{x}}{\partial \sigma^0}=0$.  Moreover, we shall assume a closed circular loop ansatz in the $xy$ plane: 
\begin{equation}
(x(t),y(t),z(t))= R(t)(\cos \theta, \sin \theta,0)\,.
\end{equation}
In a flat FLRW spacetime with scale factor $a(t)$, we get the equations:
\begin{align}
    \frac{\dot{\epsilon}}{\epsilon} &=H-a^2 \dot{R}^2\left(2 H+\frac{\dot{\mu}}{\mu}\right),
\label{eq:epsilonEOM}\\
\ddot{R}+ H \dot{R}\,+& \frac{R}{\epsilon^2}+\left(2 H+\frac{\dot{\mu}}{\mu}\right )(1-a^2 \dot{R}^2)\dot{R}=0\,,
\label{eq:REOM}
\end{align}
where $\epsilon\equiv a |R| / \sqrt{1-a^2 \dot{R}^2}$ parametrizes the physical radius of the string at the oscillation turning points and is related to its (linear) energy density \cite{Kibble:1976sj}.

We will be interested in situations where the string tension depends on time through a dynamical scalar field $\phi$ as:
\begin{equation}
\label{eq:exponential}
\mu(t)=\mu_0\,e^{-\sqrt{6}\beta \, \frac{\phi(t)}{M_p}}\ .
\end{equation}
We will first employ the techniques of \cite{SanchezGonzalez:2025uco} to study the cosmological evolution of a system including a population of cosmic superstrings and a modulus rolling towards the minimum of its potential, with and without radiation in generality. Then we will study what happens when the field stabilises at its minimum and the tension becomes oscillating, analysing the evolution of a single isolated string and whether it has any peculiar dynamics.

\section{Evolution towards the minimum} \label{sec:evolution}

All previous works on the evolution of a dilute gas of cosmic string loops assumed an exponential potential \cite{Conlon:2024uob, Brunelli:2025ems,SanchezGonzalez:2025uco}. However this approximation holds only in a portion of a phenomenologically realistic potential. In particular, when the rolling field is the volume modulus of string compactifications, it must settle into a minimum at the end of its excursion to avoid a decompactification limit. The precise structure of the potential thus becomes relevant for assessing how the field reaches its minimum and whether it settles there without overshooting. 

The potential we will focus on is the one of type IIB LVS models \cite{Balasubramanian:2005zx, Cicoli:2008va}. This class of models features (at least) two moduli fields: one controlling the overall internal volume, and the other parametrising the size of a small blow-up cycle. At leading order, the LVS potential receives three types of contributions: $\O(\alpha'^3)$ corrections to the \Kahler potential \cite{Becker:2002nn}, non-perturbative effects in the superpotential \cite{Blumenhagen:2009qh}, and a model-dependent uplift term \cite{Kachru:2003aw,Cicoli:2015ylx,Cicoli:2012fh}. After integrating out the blow-up mode, the potential depends only on the volume mode $\V$. The canonical normalisation of the volume is:
\begin{equation}
\label{eq:canonical volume}
\frac{\Phi}{M_p} = \sqrt{\frac{2}{3}} \ln \V\,.
\end{equation}
The potential for the canonically normalised volume turns out to have the following form (setting $M_p = 1$):
\begin{equation}
\label{eq:LVS potential}
V_{\rm LVS} (\Phi) = V_0 \left[(1-\epsilon \, \Phi^{3/2})\,e^{-3 \sqrt{\frac{3}{2}}\, \Phi} + \delta\,  e^{-2 \sqrt{\frac{3}{2}}\, \Phi}\right],
\end{equation}
where the monomial term comes from integrating out the blow-up mode. This potential features a steep exponential behaviour on the left of a Minkowski (or de Sitter) minimum, followed by a relatively shallow maximum. This is at the core of the overshoot problem \cite{Brustein:1992nk, Kaloper:1991mq, Barreiro:1998aj, Brustein:2004jp, Kaloper:2004yj, Battefeld:2005av, Conlon:2008cj, Acharya:2008bk, Conlon:2022pnx,Mosny:2025cyd}.

If the field reaches its minimum with too much kinetic energy, as would happen in a kination scenario, it would cross the maximum and inevitably roll towards infinity. This problem is therefore potentially present in all models with a kination epoch. However, as noted in \cite{Copeland:1997et}, the kinating fixed point of the dynamical system of a scalar field with an exponential potential is unstable.  
In fact, if even a small percentage of energy density is in any background fluid, the system will move away from the kinating fixed point towards a scaling or tracker solution.
Depending on the initial fraction of energy in the background fluid, the friction may decelerate the field enough to make it stop before it reaches the maximum. This is the typical solution to the overshoot problem, which works well for a field excursion of order $\Delta \phi \sim  \O(10)\,  M_p$ in the presence of a radiation background. The situation changes when we add a fluid of string loops. In fact, as noted in \cite{SanchezGonzalez:2025uco}, this fluid redshifts as:
\begin{equation}
\label{eq:energy density of loops}
    \rho_{\rm loop} \sim a^{-3} \, \mu^{1/2}\,.
\end{equation}
For an F-string with $\mu \sim\V^{-1}$ and a kinating volume such that $\V \sim a^3$, we get:
\begin{equation}
\rho_{\rm loop} \sim a^{-9/2}\,,
\end{equation}
i.e. the F-strings redshift \textit{faster} than radiation, making up  a rather peculiar fluid. 
However, the energy density in string loops depends on the behaviour of the scalar field through the field-dependence of the tension. Therefore, to understand how the loops influence the dynamics of the rolling modulus and vice-versa, a complete study of the dynamical system must be performed, including the contribution of the fluid of string loops. 
We shall perform this analysis combining the results of \cite{SanchezGonzalez:2025uco} with the requirement that the field does not overshoot its minimum at the end of its evolution.
We shall study the system with and without radiation.

\subsection{The dynamical system} \label{sec:dynamical system}

The Friedman equation of a system formed by a scalar field $\phi$ with potential $V(\phi)$, loops of cosmic strings, and a perfect fluid with energy density $\rho_{\rm f}$, can be written as (reintroducing $M_p$ to give a sense of the units):
\begin{equation}
\label{eq:friedmann}
3 H^2 M_p^2 =  \rho_{\rm tot} = \frac{1}{2}\dot \phi^2 + V(\phi) + \rho_{\rm loop} (\phi) + \rho_{\rm f}\,.
\end{equation}
We then have the Klein-Gordon equation for the scalar field with the loops of cosmic strings acting as a source, and the continuity equation of the background:
\begin{align}
\ddot \phi + 3 H \dot \phi + V,_\phi + \rho_{{\rm loop} ,\phi}= 0\,, \label{eq:cont eq phi}\\
\dot \rho_{\rm f} + 3H (\rho_{\rm f}+ p_{\rm f}) = 0\,.
\label{eq:cont eq fluid}
\end{align}
On the other hand, the continuity equation for the fluid of string loops reads:
\begin{equation}
\label{eq:cont eq loops}
\dot \rho_{\rm loop} + 3H \rho_{\rm loop} -\rho_{{\rm loop} , \phi} \,\dot \phi = 0\,.
\end{equation}
Differentiating \eqref{eq:friedmann} with respect to time, using the continuity equations and supposing the fluid is barotropic, $p_{\rm f} = \omega \, \rho_{\rm f}$, we get:
\begin{equation}
\label{eq:H dot}
-2 \dot H M_p^2 = \dot \phi^2 + \rho_{\rm loop} + (\omega+1)\rho_{\rm f}\,.
\end{equation}
Lets us now define the following phase space variables:
\begin{align}
X^2 & \equiv \frac{\dot \phi^2}{2 \rho_{\rm tot}}\equiv \Omega_{\rm k}\,, \label{eq:X}\\
Y^2 & \equiv \frac{V}{ \rho_{\rm tot}}\equiv \Omega_V\,, \label{eq:Y}\\
Z^2 & \equiv \frac{\rho_{\rm loop}}{ \rho_{\rm tot}} \equiv \Omega_{\rm loop}\,, \label{eq:Z}\\
W^2 & \equiv \frac{\rho_{\rm f}}{ \rho_{\rm tot}}\equiv \Omega_{\rm f}\,, \label{eq:W} 
\end{align}
where $\Omega_i$ is the fraction of energy density of the universe in the $i$-th fluid. The Friedmann equation thus reduces to the constraint:
\begin{equation}
\label{eq:Friedmann ps}
X^2+Y^2+Z^2+W^2 = 1\,,
\end{equation}
which defines the phase space of the system. 
Let us now analyse separately the cases with and without a background fluid that behaves as radiation. 

\subsection{Case without radiation}
\label{sec:no rad}

Let us first consider the case with no background fluid, setting $\rho_{\rm f} \equiv 0$ identically. In the absence of radiation, we have that:
\begin{equation}
\label{eq:Z^2 no rad}
Z^2 = 1- X^2-Y^2\,.
\end{equation}
Therefore, using the parametrisation \eqref{eq:exponential} for the tension, we can write the autonomous system in the phase space variables $X$ and $Y$. To do that, we define:
\begin{equation}
\label{eq:lambda(phi)}
\lambda(\phi) \equiv - \sqrt{\frac{2}{3}}\frac{V,_\phi}{V}\,,
\end{equation}
so that the equations of the system simplify to:
\begin{align}
 X' & = \frac{3}{2}[X(X^2-Y^2-1)+\beta (1-X^2)+ (\lambda-\beta)Y^2]\,, \label{eq:X' no fluid}\\
 Y' & = \frac{3}{2}Y(X^2-Y^2-\lambda X +1)\,, \label{eq:Y' no fluid}
\end{align}
where derivatives are taken with respect to the number of e-folds $N = \ln a$. 

The fixed points of this system for an exponential potential (i.e. $\lambda = \text{const}$) have been classified in \cite{SanchezGonzalez:2025uco}. Here we report them in Tab. \ref{tab:fixed points no rad}.
\begin{table*}[t!] 
\centering
\begin{tabular}{| m{1cm} | m{1.5cm}| m{2.3cm} |m{2.5cm}| m{7cm}|}
\hline
\centering \textbf{FP} &\centering \textbf{X} & \centering \textbf{Y} & \centering \textbf{Z} &  $\qquad \qquad \qquad \; $  \textbf{Existence} \\
\hline
\centering $\K_+$ & \centering 1 & \centering 0 & \centering 0 & $\qquad \qquad \qquad \quad \forall  \lambda\,$ and $\,\forall \beta$  \\
\hline
\centering $\K_-$ & \centering  -1 & \centering 0 & \centering 0 & $\qquad \qquad \qquad \quad \forall  \lambda\,$ and $\,\forall \beta$  \\
\hline
\centering $\mathcal{M}$  & \centering $\frac{\lambda}{2}$ &  \centering $\sqrt{1-\frac{\lambda^2}{4}}$ & \centering 0  & $\qquad \qquad \qquad \qquad   \forall \beta,\, \lambda \leq 2$ \\
\hline
\centering $\mathcal{L}$ &\centering  $\beta$ &\centering  0 & \centering $\sqrt{1-\beta^2}$   & $\qquad \qquad \qquad \qquad  \forall \lambda, \, \beta  \leq 1$  \\
\hline 
\centering $
\mathcal{T}_1$ & \centering  $\frac{1}{\lambda - \beta}$ & \centering  $\frac{\sqrt{\beta^2 +1 - \lambda\beta }}{\lambda-\beta}$ & \centering $\frac{\sqrt{\lambda^2 -2 - \lambda\beta }}{\lambda-\beta}$ & $\quad \beta \leq 1 \;$ and $\; \frac{\beta}{2} + \sqrt{\frac{\beta^2}{4}+2} \leq \lambda \leq \beta + \frac{1}{\beta}$ \\
\hline
\end{tabular}
\caption{Fixed points (FP) of the autonomous system \eqref{eq:X' no fluid}-\eqref{eq:Y' no fluid} with existence conditions.}
\label{tab:fixed points no rad}
\end{table*}
$\mathcal{K}_{\pm}$ are the
\textit{kinating} fixed points where the total energy density is stored in the kinetic energy of the rolling field. $\mathcal M$ is instead the \textit{field domination} fixed point where the energy density of the universe is dominated by the one of the scalar field. $\mathcal{L}$ is the \textit{loop tracker} where the kinetic energy of the field tracks the energy density of the string loops $\rho_{\rm loop}$. Finally, $\mathcal{T}_1$ is a \textit{mixed tracker} where the energy density is distributed in fixed ratios between kinetic energy, potential energy and cosmic string loops.

As previously stated, in the absence of any external fluid, a field rolling down a steep potential towards a shallow minimum would overshoot it.\footnote{Let $\Phi_{\rm max}$ be the field value where the potential has its local maximum, and let $\Phi_*<\Phi_{\rm max}$ be such that $V(\Phi_*) = V(\Phi_{\rm max})$. Then the previous statement is true for all $\Phi_0 < \Phi_*$.} Let us now analyse how this situation changes in the presence of a fluid of cosmic string loops.

To better understand the overall dynamics of the system and its late-time behaviour, we shall perform a stability study of its fixed points to assess the final attractors for general $\beta$ and $\lambda$ in the case of an exponential potential. The stability map  is provided in Fig. \ref{fig:stab_no_fluid} and the details of the stability regions are presented in the Appendix.
\begin{figure}[h!]
\centering
\includegraphics[width=0.7\textwidth]{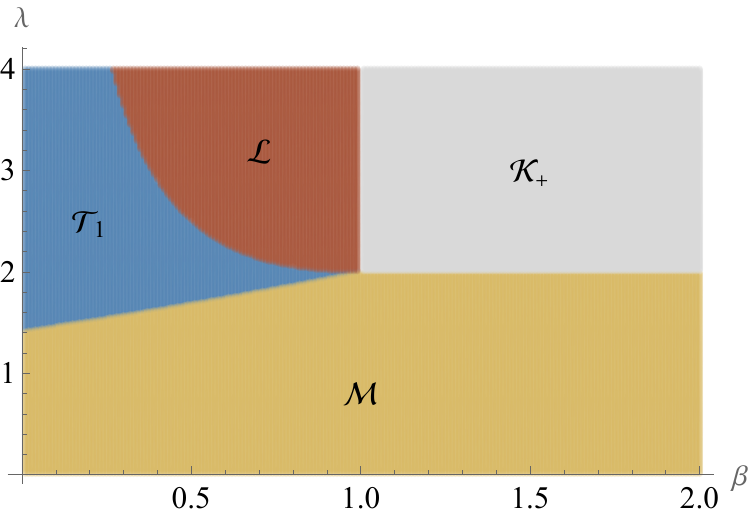}
\caption{Stability map of the fixed points of the dynamical system \eqref{eq:X' no fluid}-\eqref{eq:Y' no fluid} with exponential potential.}
\label{fig:stab_no_fluid}
\end{figure}

\begin{figure} [h!]
    \centering
    \includegraphics[width=0.7\textwidth]{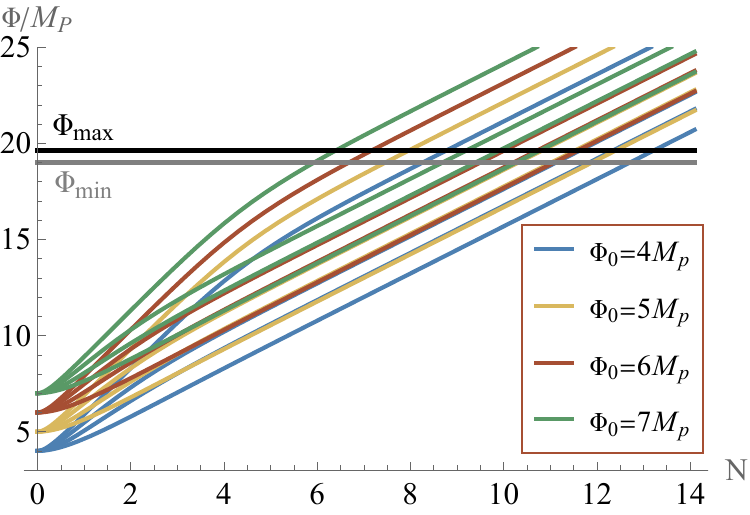}
    \caption{Evolution of $\Phi$ towards the minimum of the potential \eqref{eq:LVS potential} located at $\Phi_{\rm min} = 19 M_p$. For each initial value $\Phi_0$, 4 initial densities of loops of F-strings with $\beta = 1/2$ have been tested: $Z_0^2 =\{ 0.01, 0.51, 0.84, 0.99\}$. All of them overshoot the minimum and the maximum (grey and black lines respectively).}
    \label{fig:overshoot}
\end{figure}

If we consider the full LVS potential \eqref{eq:LVS potential}, we lose the general validity of this map. However, for small values of the field $\Phi$, sufficiently far from the minimum $\Phi_{\rm min}$ where all the contributions compete, the potential can still be reliably approximated by an exponential:
\begin{equation}
\label{eq:Lvs exponential}
V \simeq V_0 \, e^{-3\sqrt{\frac{3}{2}}\, \Phi}\,,
\end{equation}
which gives $\lambda = 3$. Now, depending on the type of strings in the fluid of cosmic string loops, the behaviour of the system is different. For F-strings, the tension is simply given by the string scale which, in Einstein frame, can be expressed as:
\begin{equation}
\label{eq:F-string tension}
\mu \sim M_s^2 = \frac{\sqrt{g_s} \,M_p^2}{4 \pi \V}\,.
\end{equation}
Canonically normalising the volume as in \eqref{eq:canonical volume}, the tension becomes exponential in $\Phi$:
\begin{equation}
\mu \sim M_p^2 \, e^{- \sqrt{\frac{3}{2}}\,\frac{\Phi}{M_p}}\,.
\end{equation}
Comparing this expression with the general parametrisation of the tension given in (\ref{eq:exponential}), we find:
\begin{equation}
\label{eq:beta f-string}
\beta = 1/2\,.
\end{equation}
If the field range for which the approximation \eqref{eq:Lvs exponential} holds is sufficiently large, the system will evolve towards its final attractor which corresponds to the loop tracker $\mathcal{L}$, as can be seen from Fig. \ref{fig:stab_no_fluid} with $\beta = 1/2$ and $ \lambda  = 3$. This is one of the main results of \cite{SanchezGonzalez:2025uco} which also found that in this scenario the loop energy density fraction grows significantly up to about $75\%$. However, as can be seen in Tab. \ref{tab:fixed points no rad}, the kinetic energy of the field for $\beta = 1/2$ is still $25\%$ of the total energy density, and so the rolling volume would generally overshoot the minimum. This is shown in Fig. \ref{fig:overshoot}. Here, the field $\Phi$ rolls down the full potential \eqref{eq:LVS potential} in a background of loops of cosmic F-strings with various initial energy densities $Z_0^2\equiv\Omega_{\rm loop}^{(0)}$. In (\ref{eq:LVS potential}) we have chosen $V_0=1$, $\epsilon=0.013$ and $\delta\simeq 5.397 \times 10^{-12}$ to obtain a Minkowski vacuum at $\Phi_{\rm min}=19\,M_p$ that would correspond, for $W_0\sim\mathcal{O}(10)$, to masses of the volume mode just above the bound of the cosmological moduli problem, $m_{\mathcal{V}}\sim W_0 M_p/\mathcal{V}^{3/2}\gtrsim \mathcal{O}(30)$ TeV. It can be seen that the field overshoots the maximum of the potential, if released with vanishing velocity for a wide range of initial conditions. 

Let us now investigate what happens when the fluid of cosmic string loops is made of \emph{effective strings}. These are the result in the EFT of $p$-branes wrapping $(p-1)$-cycles in the internal dimensions, leaving one extended space dimension that looks like a string in spacetime.\footnote{In \cite{Brunelli:2025ems} these objects were called \textit{EFT strings.}} As shown in \cite{Revello:2024gwa, Brunelli:2025ems}, the tension of such strings acquires an additional dependence on the volume of the wrapped cycle. In particular, if the wrapped $(p-1)$-cycle $\Sigma_{p-1}$ has volume $\text{Vol}(\Sigma_{p-1})$, then:
\begin{equation}
\label{eq:EFT tension}
\mu \sim M_s^{p+1} \, \text{Vol}(\Sigma_{p-1})\,.
\end{equation}
For a simple volume of the form $\V = \tau^{3/2}$ there are two possibilities:
\begin{enumerate}
\item \textit{D3-branes} wrapped around a 2-cycle (henceforth referred to as \textit{D3-strings}): In this case, $p=3$ and $\text{Vol}(\Sigma_2) \simeq \V^{1/3} M_s^{-2}$, and so the tension of the string would go as:
\begin{equation}
\mu \sim M_s^2 \,\V^{1/3}\sim M_p^2\,\V^{-2/3}\sim e^{-\sqrt{2/3} \,\Phi/M_p}\,.
\end{equation}
Comparing with (\ref{eq:exponential}) we find:
\begin{equation}
\beta = 1/3\,.
\end{equation}
The point $(\beta = 1/3, \lambda = 3)$ lies in the blue region of the of parameter space shown in Fig. \ref{fig:stab_no_fluid}, implying that the final attractor of the system is the mixed tracker $\mathcal{T}_1$. However, the convergence towards $\mathcal{T}_1$ turns out to be rather slow.

\item \textit{NS5-branes} wrapped around a 4-cycle (henceforth referred to as \textit{NS5-strings}): In this case, $p=5$ and $\text{Vol}(\Sigma_4) \simeq \V^{2/3} M_s^{-4}$, and so the tension of the string would go as:
\begin{equation}
\mu \sim M_s^2 \V^{2/3} \sim M_p^2\,\V^{-1/3}\sim e^{-1/\sqrt{6} \, \Phi/M_p}\,.
\end{equation}
Comparing again with (\ref{eq:exponential}), we get:
\begin{equation}
\beta = 1/6\,.
\end{equation}
The point $(\beta = 1/6, \lambda = 3)$ is again in the blue region of the parameter space shown in Fig. \ref{fig:stab_no_fluid}, implying that the final attractor of the system is again $\mathcal{T}_1$. Here, it turns out that the evolution towards the final attractor is faster, but in absolute terms it is still rather slow. 
\end{enumerate}

In both cases listed above, the final attractor of the system is no longer $\mathcal L$ but $\mathcal T_1$. The loop tracker fixed point is a saddle and for our choice of initial conditions, the system still tends to evolve towards and to linger around it for quite a long time before heading towards the true attractor $\mathcal T_1$.
For any realistic value of the initial condition $\Phi_0$ that guarantees control over the EFT expansion, the field reaches its minimum \textit{before} the system has moved towards $\mathcal T_1$. Therefore, the field will overshoot its minimum if the kinetic energy $X^2$ at $\mathcal L$ is too high.

\begin{figure}[h!]
\centering
\includegraphics[width=0.7\textwidth]{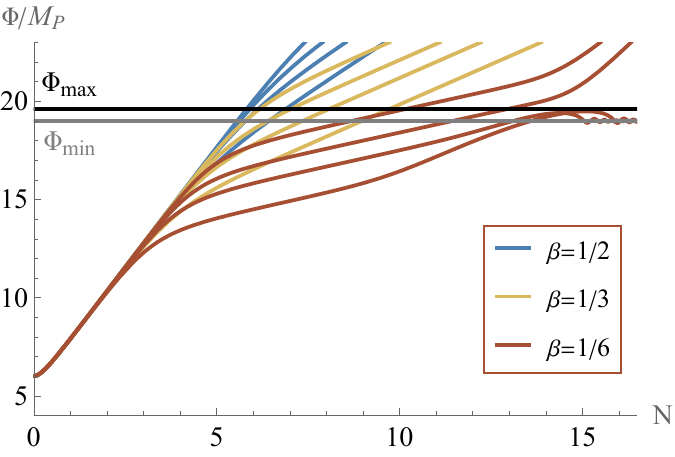}
\caption{Evolution of $\Phi$ from $\Phi_0 = 6 M_p$ towards the minimum of the potential \eqref{eq:LVS potential} located at $\Phi_{\rm min} = 19 M_p$ for the 3 values of $\beta$. For each $\beta$, 4 initial densities of loops have been tested: $Z_0^2 =\{ 10^{-4}, 4 \times 10^{-4}, 1.2 \times 10^{-3}, 6.4 \times 10^{-3}\}$. For $\beta =1/2$ and  $\beta =1/3$, all of them overshoot, while for $\beta= 1/6$ the trajectories with the higher values of $Z_0^2$ do not overshoot.}
    \label{fig:beta overshoot}
\end{figure}

In the case of D3-strings, $X^2 \simeq \beta^2 = 1/9$ is not low enough to avoid overshooting, and the system has an evolution which is qualitatively similar to the F-string case, as can be seen in Fig. \ref{fig:beta overshoot}. 
The case of NS5-strings as a loop fluid is instead quite promising for avoiding overshooting. In fact, as can be seen in Fig. \ref{fig:beta overshoot}, for a sufficiently high fraction of energy density in loops, the field does not overshoot. 

\begin{figure}[h!]
    \centering
    \begin{subfigure}{0.49\textwidth}
        \centering
        \includegraphics[width=\linewidth]{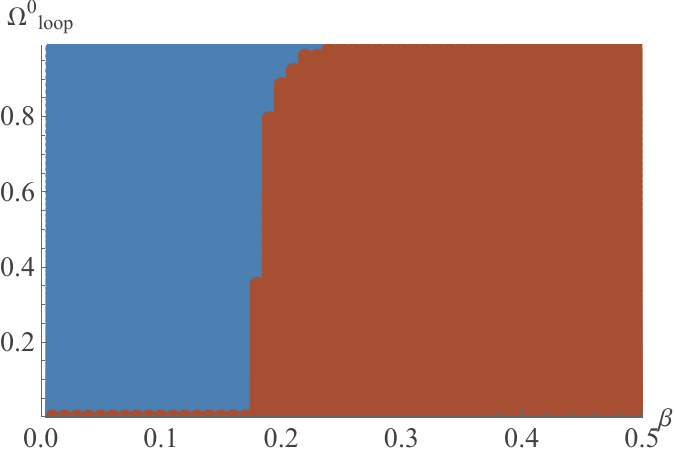}
    \end{subfigure}
    \hfill
    \begin{subfigure}{0.49\textwidth}
        \centering
        \includegraphics[width=\linewidth]{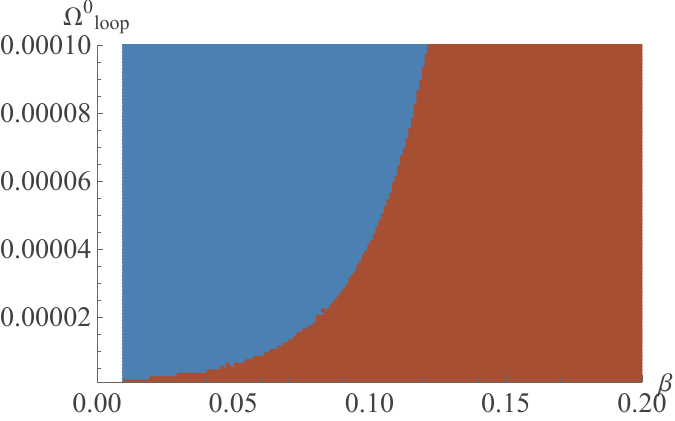}
    \end{subfigure}
    \caption{Systematic scan over the $(\beta, \Omega_{\rm loop}^{0})$ space. The blue region corresponds to values of the parameters for which the field does not overshoot its minimum while red region represents the parameters for which it does. The other initial conditions are fixed as: $\Phi_0=6$, $\dot{\Phi}_0=0$ and $\Phi_{\rm min}=19$ \textit{Left:} Scan over all the meaningful values of $\beta$, up to $\beta = 1/2$. The plot shows that the field will always overshoot for $\beta > 0.23$. \textit{Right:} Zoom over the $\beta<0.2$ region, showing how the minimal $\Omega_{\rm loop}^0$ grows exponentially with $\beta$.}
     \label{fig:overshoot scan}
\end{figure}

\begin{figure}
\centering
\includegraphics[width=0.7\textwidth]{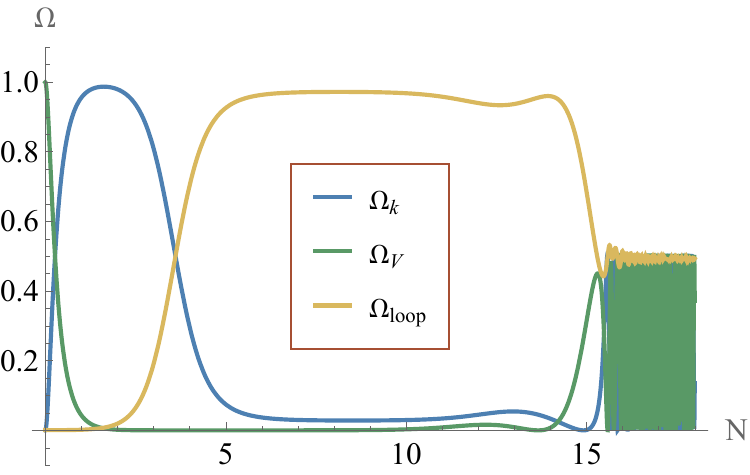}
\caption{Evolution of the fractional energy densities of the dynamical system in the case of NS5-strings with $\beta = 1/6$ and no radiation. The oscillating behaviour at late times indicates that the field settles into the minimum. The initial conditions for the field and relative energy densities are: $\Phi_0 = 6 M_p$, $X^2_0 = 0$, $Y^2_0=1-10^{-3}$, $Z_0^2 = 10^{-3}$. }
    \label{fig:NS5 no rad}
\end{figure}

In Fig. \ref{fig:overshoot scan} we perform a systematic scan over the $(\beta,\Omega_{\rm loop}^0)$ parameter space, in order to determine the values of these parameters that do not lead to volume modulus overshoot. In this analysis all other parameters are kept fixed, in particular $\Phi_0=6M_p$, $\dot{\Phi}_0=0$ and $\Phi_{\rm min}=19 M_p$. We observe that a non-vanishing energy density in string loops is always required to avoid overshooting and that this fraction increases with increasing $\beta$. Furthermore we observe that overshoot is inevitable for $\beta > 0.23$, in accordance with the findings in Fig. \ref{fig:beta overshoot}.
We can then study the evolution of the fractions of the energy densities $\Omega_{\rm k}$, $\Omega_V$, and $\Omega_{\rm loop}$ along the solutions of the autonomous system \eqref{eq:X' no fluid}-\eqref{eq:Y' no fluid} with $\beta = 1/6$ which do not overshoot. An illustrative result is shown in Fig. \ref{fig:NS5 no rad}. 
Here, the initial conditions for the energy densities have been chosen keeping in mind a scenario where the volume modulus starts its evolution right at the end of inflation, so that, as it thaws, all its energy density is originally in its potential. This can happen in hybrid inflation models as brane-antibrane inflation \cite{Burgess:2001fx, Dvali:2001fw, Kachru:2003sx, Baumann:2007ah, Baumann:2007np,Cicoli:2024bwq}, where string loops can be produced at the end of inflation when the volume acts as a waterfall field.

Note that, even for a small initial fraction of energy density in loops, the field does not overshoot. There are a few other interesting features of this case. First of all, during the evolution of the system, the fraction of energy density in loops grows significantly, reaching values that are even higher than the ones found in \cite{SanchezGonzalez:2025uco}. The reason is the lower value of $\beta$: the system approaches the loop tracker fixed point $\mathcal L$ where $\Omega_{\rm loop}  = 1-\beta^2 = 35/36 \simeq 0.972$, i.e. where the energy density in effective string loops can be as high as $97.2\%$! 

\begin{figure}
\centering
\includegraphics[width=0.7\textwidth]{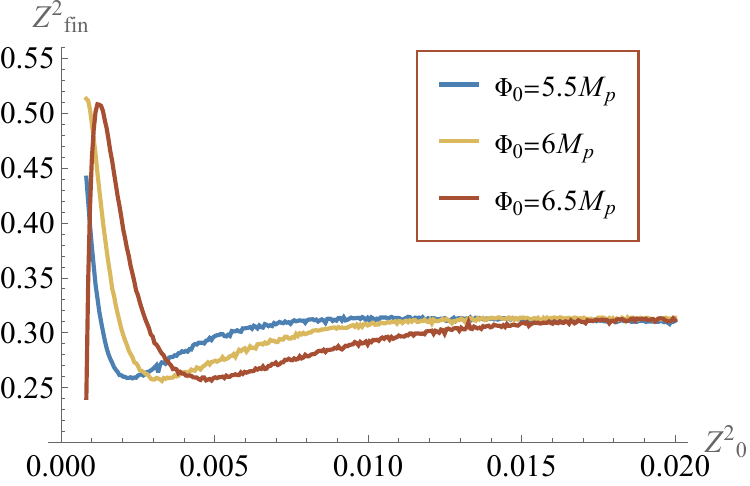}
\caption{Final averaged fraction of energy density in NS5-loops while $\Phi$ oscillates around its minimum as a function of the initial energy density of the loop fluid $Z_0$ and for different initial values of the field $\Phi_0$. For higher values of $\Phi_0$, the peak shifts to the right. For $\Phi_0 = 6M_p$, the peak is at $Z_0^2 \simeq 8.6 \times 10^{-4}$ , close to the  value chosen for Fig. \ref{fig:NS5 no rad}.}
\label{fig:final density}
\end{figure}

What happens after the field has reached its minimum and starts oscillating is perhaps even more interesting. As shown in Fig. \ref{fig:NS5 no rad}, the fraction of energy density in loops stabilises to about $50\%$! The reason for this is the following. As seen in \eqref{eq:energy density of loops}, $\rho_{\rm loop}$ redshifts as matter plus the additional factor due to the decreasing tension. However, as soon as the field reaches its minimum, the tension starts oscillating around a fixed value, i.e. it stops decreasing. This means that from this point onwards $\rho_{\rm loop} \sim a^{-3}$, which is the same scaling behaviour of a field oscillating around its minimum, $\rho_\phi \sim a^{-3}$. This implies that the fraction of the energy densities remains on average constant:
\begin{equation}
\label{eq:constant fraction no fluid}
 \frac{\Omega_{\rm loop}}{\Omega_{\rm k}+\Omega_V}\equiv\frac{Z^2}{X^2+Y^2} \simeq \text{const}\,.
\end{equation}
We expect the fraction of energy density in loops to remain constant until they decay into gravitational radiation. This is very interesting in sight of the possible spectrum of emitted gravity waves, as the NS5-loops constitute a relevant fraction of the energy density of the universe (in Fig. \ref{fig:NS5 no rad} around $50\%$) throughout the early matter domination period. It is important to stress that this fraction exhibits a non-trivial dependence on the precise initial density of strings. In Fig. \ref{fig:final density} we show how the value to which $\Omega_{\rm loop}$ converges during the oscillating phase depends on $\Omega_{\rm loop}^{(0)}\equiv Z_0^2$ for different choices of initial conditions for $\Phi$. It is interesting to notice that $\Omega_{\rm loop}$ tends to around $30\%$ for large values of $Z_0^2$, while it can get to values of $\mathcal{O}(50\%)$ for small $Z_0^2$.

\subsection{Multiple string loop species} \label{sec:multiple beta}

The above analysis concentrates on a single species of fluid at a time. It is however relevant to ask if any interesting dynamics can arise in the presence of multiple loop species. Before we attempt this analysis, let us look at how the various energy densities scale with the rolling volume modulus. From the discussion above, we have that the energy density of F-strings scales as ($M_p = 1$):
\begin{equation}
\rho_F\propto a^{-3} \mathcal{V}^{-1/2} = a^{-3} e^{-\frac{1}{2}\sqrt{\frac{3}{2}}\Phi}\,,
\end{equation}
whereas that of D3-strings:
\begin{equation}
\rho_3\propto a^{-3} \mathcal{V}^{-1/3}= a^{-3} e^{-\frac{1}{3}\sqrt{\frac{3}{2}}\Phi}\,,
\end{equation}
and for NS5-strings:
\begin{equation}
    \rho_5\propto a^{-3} \mathcal{V}^{-1/6}= a^{-3} e^{-\frac{1}{6}\sqrt{\frac{3}{2}}\Phi}.
\end{equation}
Given that $\Phi$ is growing as it rolls towards the minimum, if all three species are present at early times with similar energy densities, we see that NS5-strings will quickly become most abundant and give the dominant contribution to the cosmological dynamics. This implies that, even in this more general case, overshooting may be avoided with the system evolving through the NS5-brane loop tracker $\mathcal{L}$ before converging to the minimum.
This happens provided that we neglect the interactions among string species. If such interactions were not negligible, the dynamics of this system would become more complicated and a more refined analysis would be needed.

\subsection{Case with radiation}\label{sec:radiation}

If we include 
background fluid with equation of state 
$p_ {\rm f} = \omega \, \rho_{\rm f}$, the dynamics of the system becomes even richer. Recalling that $\Omega_{\rm f}\equiv W^2$ and including the continuity equation of the fluid in the dynamical system, we get:
\begin{align}
   X' & = \frac{3}{2}\bigg[X \left(X^2-Y^2+ \omega W^2-1\right) +\beta \left(1-X^2-W^2\right)+ (\lambda-\beta)Y^2\bigg], \label{eq:X' with fluid}\\
 Y' & = \frac{3}{2}Y\left(X^2-Y^2+ \omega W^2-\lambda X +1\right), \label{eq:Y' with fluid}\\
 W' & = \frac{3}{2}W \left[X^2-Y^2+ \omega\left(W^2-1\right)\right].\label{eq:W' with fluid}
\end{align}
\begin{table}[t!] 
    \begin{tabular}{| m{0.5cm} | m{0.7cm}| m{1cm} |m{3.7cm}| m{2.9cm} |m{6cm}|}
        \hline
        \centering \textbf{FP} &\centering \textbf{X} & \centering \textbf{Y} & \centering \textbf{Z} &  \centering \textbf{W} &$\qquad \quad $  \textbf{Existence} \\
        \hline
        \centering $\mathcal {S}$ & \centering 
        $\frac{\omega+1}{\lambda}$& \centering 
        $\frac{\sqrt{1-\omega^2}}{\lambda}$& \centering 0 & \centering 
        $\sqrt{1-\frac{2(\omega+1)}{\lambda^2}}$&$\; \lambda \geq \sqrt{2(\omega+1)}\,$, $\, \beta \neq \lambda \frac{\omega}{(\omega+1)}$  \\
        \hline
        \centering $\mathcal{LR}$ & \centering  $\frac{\omega+1}{\lambda}$ & \centering $y_{\rm lr}$ & \centering 
        $\sqrt{\frac{(1-\omega^2)(1+\omega)}{\omega \lambda^2}-\frac{\omega+1}{\omega}y^2_{\rm lr}}$&
        \centering 
        $\sqrt{1-\frac{(\omega+1)^2}{\omega\lambda^2}+\frac{ y_{\rm lr}^2}{\omega}}$&
        $ \lambda \geq \sqrt{2(\omega+1)}$, $ \, \beta = \lambda \frac{\omega}{(\omega+1)},\,  \omega > 0$ \\
       \hline
      \centering $\mathcal{T}_2$  & \centering
      $\frac{\omega}{\beta}$ &  \centering 0& \centering 
      $\frac{\sqrt{\omega(1-\omega)}}{\beta}$& \centering $\sqrt{1-\frac{\omega}{\beta^2}}$ & $\qquad \qquad  \forall \lambda,\, \beta \geq \sqrt{\omega}$ \\
      \hline
       \centering $\mathcal{F}$ &\centering  0 &\centering  0 & \centering 0 & \centering 1 & $\qquad \qquad \,   \forall \lambda$ and $\,\forall \beta$  \\
        \hline
    \end{tabular}
    \caption{Fixed points (FP) of the autonomous system \eqref{eq:X' with fluid}-\eqref{eq:Y' with fluid}-\eqref{eq:W' with fluid} with existence conditions. These all exist together with the fixed points of Tab. \ref{tab:fixed points no rad}.}
    \label{tab:fixed points with rad}
\end{table}
Once again, the fixed points for this system in the case of an exponential potential, i.e. $\lambda=\text{const}$ in the case of a radiation fluid $\omega = 1/3$, have been catalogued in \cite{SanchezGonzalez:2025uco}, and we report them in Tab. \ref{tab:fixed points with rad}.
All the fixed points of the dynamical system in the absence of  fluid \eqref{eq:X' no fluid}-\eqref{eq:Y' no fluid}  are still fixed points of this system, as they belong to the $W=0$ section of the phase space, albeit with altered stability properties. The new fixed points of Tab. \ref{tab:fixed points with rad} are exclusive to the case with radiation. In fact, we have the \textit{fluid dominated} fixed point $\mathcal F$ in which all of the energy density of the universe is in radiation. In the \textit{scaling} solution $\mathcal S$, or \textit{ tracker}, also analysed in \cite{Copeland:1997et, Brunelli:2025ems}, the loops do not play any role. Then we have two new tracker solutions, namely the \textit{loop-radiation tracker} $\mathcal{LR}$ and a second \textit{mixed tracker} $\mathcal{T}_2$.

The fixed point $\mathcal{LR}$ exists only in the specific case where $\beta = \lambda \omega/(\omega+1)$, and is parametrised by its $Y$ coordinate, $y_{\rm lr}$. 
In fact, this is not a single fixed point but a line in phase space with the $Y$ coordinate lying in the range $y_{\rm min} \leq y_{\rm lr}\leq \frac{\sqrt{1-\omega^2}}{\lambda }$, with $y_{\rm min}$ depending on $\lambda$ as:
\begin{equation}
y_{\rm min} = \begin{cases}
        \sqrt{\frac{(\omega+1)^2}{ \lambda^2}-\omega} \quad & \text{for } \, \sqrt{2(\omega+1)}\leq \lambda \leq \frac{\omega+1}{\sqrt \omega}\\
        0 \quad & \text{for }\, \lambda \geq \frac{\omega+1}{\sqrt \omega}
    \end{cases}\,.
\end{equation}
We note that this fixed point plays no role in the following discussion since we lack a concrete realisation of effective strings with $\beta=3/4$ (for $\lambda=3$, $\omega = 1/3$).

At the fixed point $\mathcal{T}_2$ the energy densities are distributed in fixed fractions, this time completely depleting the potential energy. 

From this point onward, we focus on the case where the background fluid is radiation, $\omega= 1/3$, as this is the most relevant case.

Depending on the choice of initial conditions, we find that the presence of radiation can play the known role to avoid overshooting, even for F-strings. If the initial fraction of energy density in radiation is sufficiently large with respect to the initial density of loops, the system can approach the radiation tracker $\mathcal S$ where the field can be sufficiently slowed down to avoid overshooting. To assess this more precisely, let us first study the stability of the fixed points of this new system with respect to the parameters $\beta$ and $\lambda$, as we did for the case without radiation. The stability chart is shown in Fig. \ref{fig:stability rad}, with more details given in the Appendix.

\begin{figure}
\centering
\includegraphics[width=0.7\textwidth]{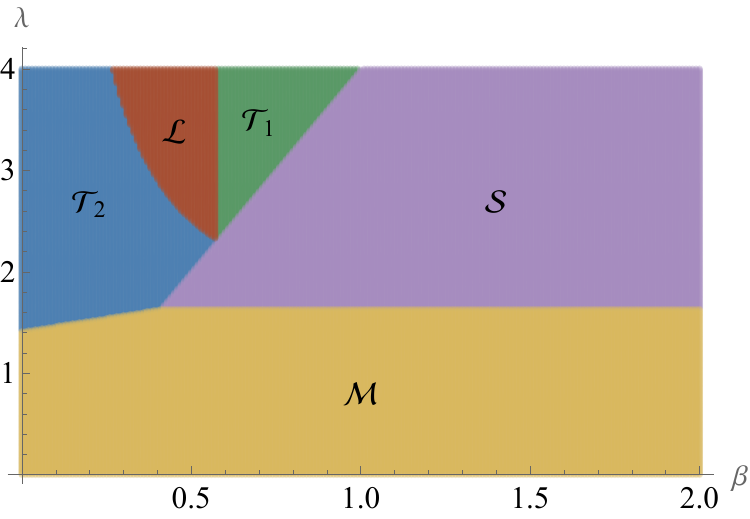}
\caption{Stability map of the fixed points of the dynamical system \eqref{eq:X' with fluid}-\eqref{eq:W' with fluid} for an exponential potential and a radiation background fluid.}
\label{fig:stability rad}
\end{figure}

\begin{figure}[h!]
\centering
\includegraphics[width=0.7\textwidth]{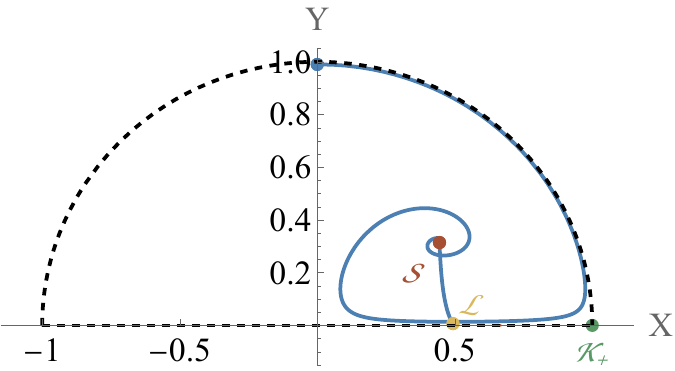}
\caption{Evolution in the $(X,Y)$-plane projection of the phase space of the dynamical system with an exponential potential with $\lambda = 3$. The system evolves first towards $\mathcal K_+$, then spirals around the radiation tracker $\mathcal S$, and eventually evolves towards the attractor  $\mathcal L$. The initial conditions are: $X_0^2 = 0$, $Y^2_0 = 0.9898$,  $Z_0^2 = 10^{-4}$, $W_0^2 = 0.01$.}
\label{fig:exponential ds}
\end{figure}

The exponential approximation \eqref{eq:Lvs exponential} of the LVS potential in the runaway region, in the case of F-strings, corresponds to the point $(\beta = 1/2, \lambda = 3)$ in parameter space, yielding the loop tracker $\mathcal{L}$ as the final attractor, in agreement with the results of \cite{SanchezGonzalez:2025uco}. However, if the field reaches the minimum \textit{before} the system approaches $\mathcal L$, there is a chance of avoiding overshooting. In fact, for a purely exponential potential without a minimum, the system will first hover around the radiation tracker fixed point $\mathcal S$ and then evolve towards the final attractor $\mathcal L$, as can be appreciated in Fig. \ref{fig:exponential ds}.

The radiation tracker can prevent the field from overshooting. However, as can be seen from Tab. \ref{tab:fixed points with rad}, the radiation tracker has no energy density in loops. $Z^2$ increases as the system moves from $\mathcal S$ to $\mathcal L$, and reaches its maximum at the attractor. If the field reaches its minimum when the system is just moving from $\mathcal S$ to $\mathcal L$, then the cosmic F-string loops can still hold a non-negligible fraction of energy density. 

\begin{figure}
\centering
\includegraphics[width=0.7\textwidth]{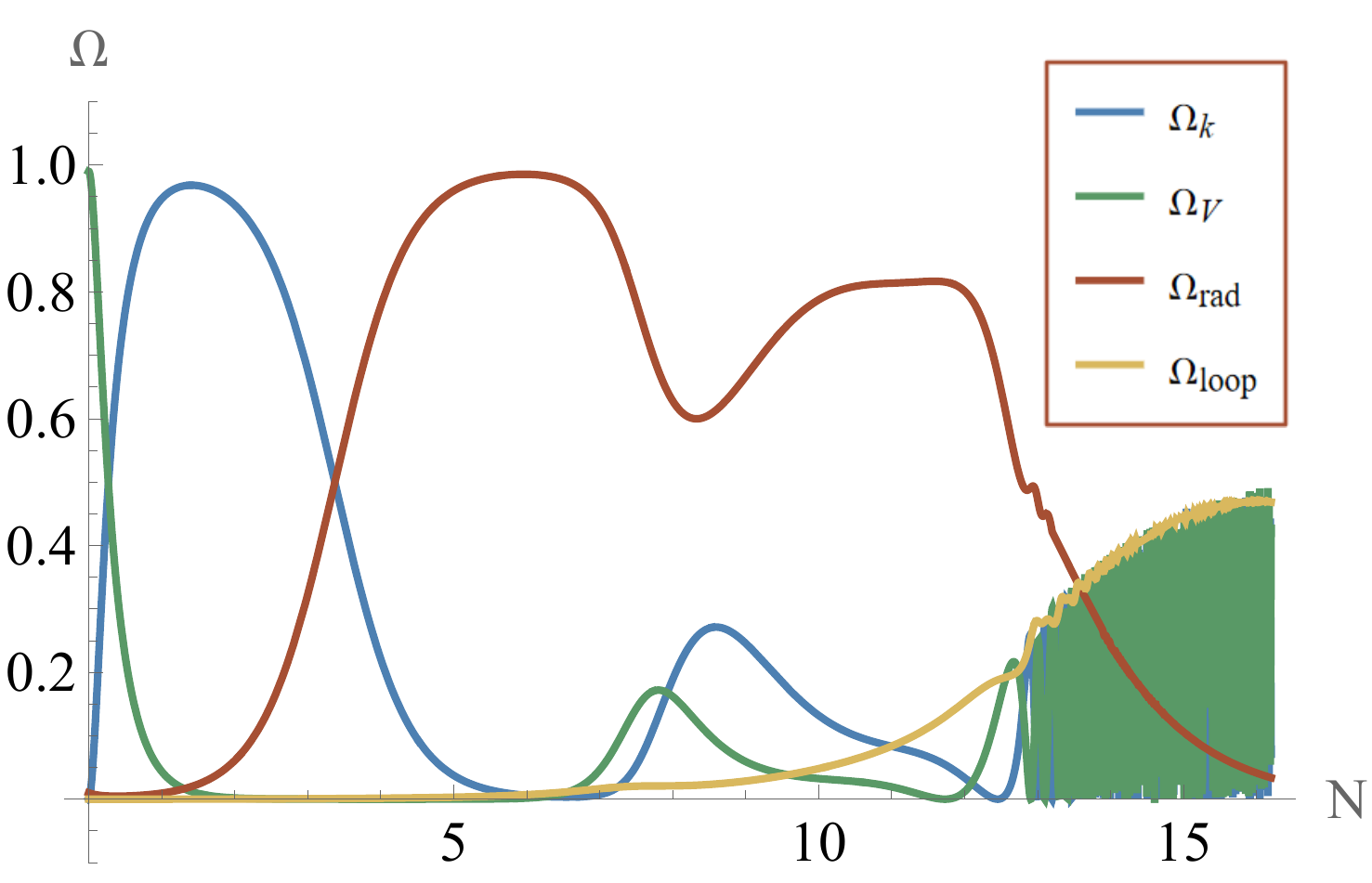}
\caption{Evolution of the fractional energy densities of the dynamical system with F-strings ($\beta = 1/2$) and radiation. The field does not overshoot and $\Omega_{\rm loop}$ keeps growing after the field reaches the minimum. The initial conditions are: $\Phi_0 = 6 M_p$, $X^2_0 = 0$,  $Z_0^2 = 5 \times 10^{-5}$, $W_0^2 = 10^{-2}$, $Y^2_0=1-X_0^2-Z_0^2-W_0^2 = 0.98995$.}
    \label{fig:F string rad}
\end{figure}

However, if the initial fraction of energy density in F-string loops ($\beta = 1/2$) is too large, the system will evolve rapidly towards $\mathcal L$, making overshooting inevitable. On the other hand, if this fraction is too small, the loops will remain a negligible fraction of the energy density before the field reaches its minimum, and the system reduces to a field rolling in a radiation background. This translates into the requirement that there must be a hierarchy between $W_0^2$ and $Z_0^2$ to prevent overshooting. The evolution of the energy density fractions of the system with radiation is displayed in Fig.  \ref{fig:F string rad} which shows that the field does not overshoot.

As shown in Fig. \ref{fig:exponential ds}, the system evolves first towards a kination epoch, then lingers around the radiation tracker for a few efolds and finally reaches the minimum as it is evolving towards $\mathcal L$. Thus $\Omega_{\rm loop}=Z^2$ starts growing before the field starts oscillating, but without overcoming the dominant radiation. An interesting feature of this system emerges as $\Phi$ starts oscillating around its minimum. At this stage, the cosmic string loops behave (on average) as matter and satisfy \eqref{eq:constant fraction no fluid}. However, now the system includes also radiation which redshifts \textit{faster} than matter, as $\rho_{\rm rad} \sim a^{-4}$. Therefore, the ratios $Z^2/W^2$ and $(X^2+Y^2)/W^2$ both grow during the oscillations. This implies that, as $W^2$ decreases, $X^2+Y^2$ increases, so does $Z^2$ in order for \eqref{eq:constant fraction no fluid} to be always satisfied. The resulting effect is a growth in the fraction of the energy density in loops as the the volume modulus oscillates around minimum. Just like in the case with no radiation, $\Omega_{\rm loop}=Z^2$ can reach significant values, around $50\%$, as can be seen from Fig. \ref{fig:F string rad}. As radiation redshifts away, this fraction settles and we are back to the case without radiation. 

\begin{figure}
\centering
\includegraphics[width=0.7\textwidth]{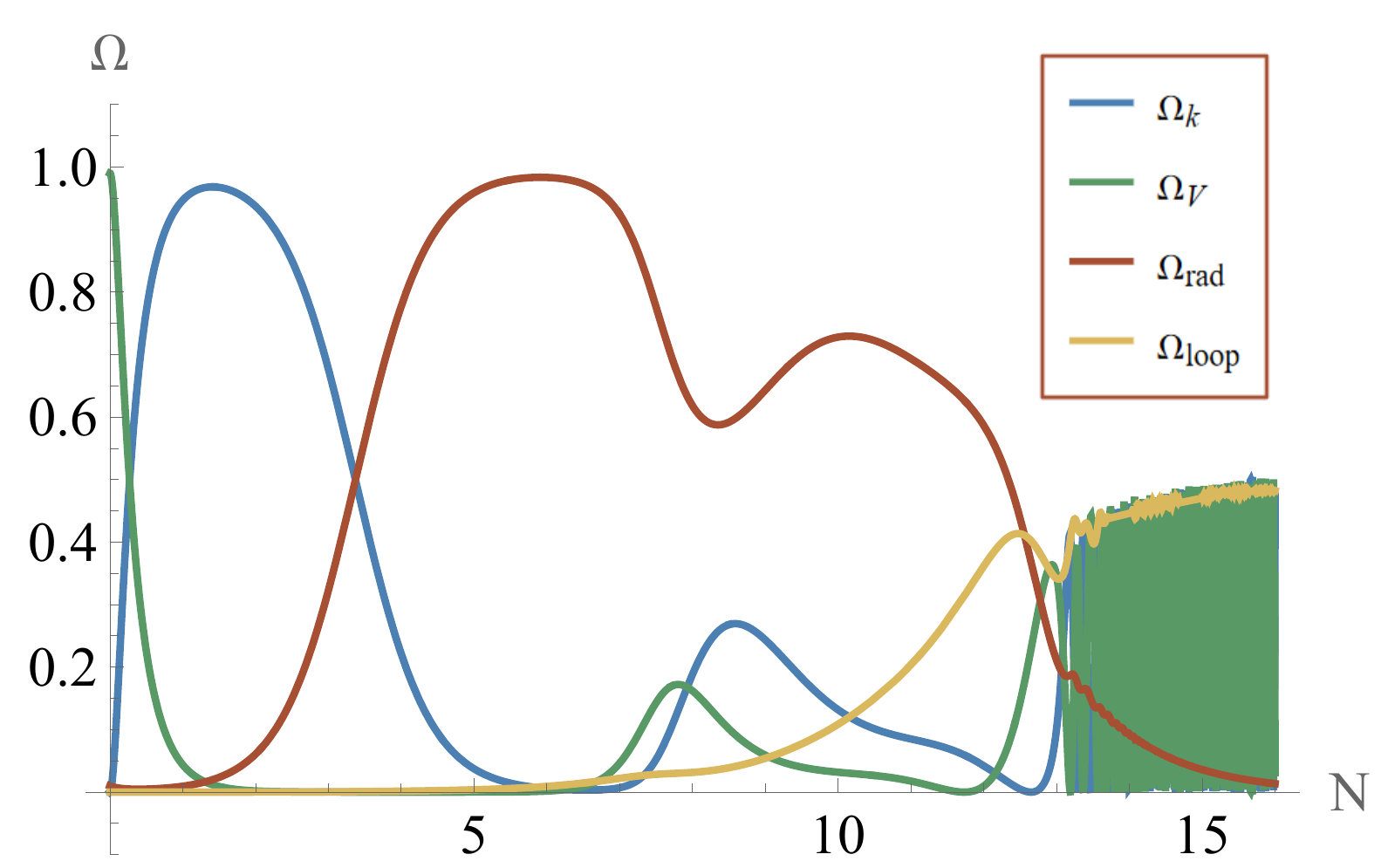}
\caption{Evolution of the fractional energy densities of the dynamical system with D3-strings ($\beta = 1/3$) and radiation. The field does not overshoot and $\Omega_{\rm loop}$ keeps growing after the field reaches the minimum. The initial conditions are: $\Phi_0 = 6 M_p$, $X^2_0 = 0$,  $Z_0^2 = 1 \times 10^{-5}$, $W_0^2 = 10^{-2}$, $Y^2_0=1-X_0^2-Z_0^2-W_0^2 = 0.98999$.}
    \label{fig:D3 rad}
\end{figure}

A similar behaviour is expected for effective strings. However, as argued in Sec. \ref{sec:multiple beta}, for $\beta <1/2$ the energy density of loops redshifts slower with respect to the F-string case, and so the loop tracker is reached faster. Thus, for the initial fraction of energy densities used to plot Fig. \ref{fig:F string rad}, D3-strings with $\beta = 1/3$ will lead the field to overshoot. However, if the hierarchy between $W_0^2$ and $Z_0^2$ is higher, the system will mimic precisely what happens in Fig. \ref{fig:F string rad}. The evolution of the energy densities in a case with $\beta=1/3$ where the field does not overshoot is shown in Fig. \ref{fig:D3 rad}. 

\begin{figure}[h!]
    \centering
    \includegraphics[width=0.6\textwidth]{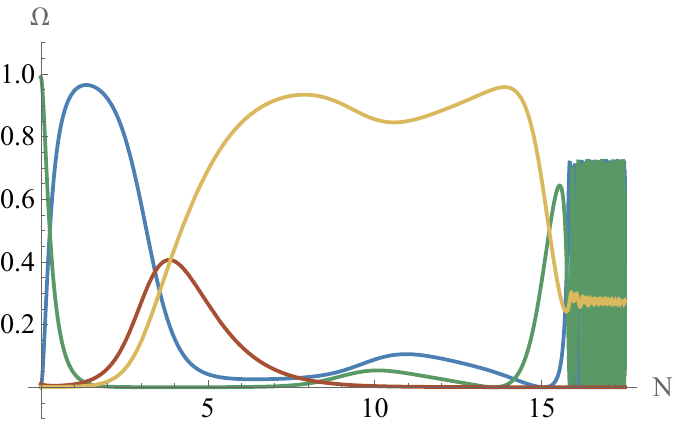}
    \includegraphics[width=0.6\textwidth]{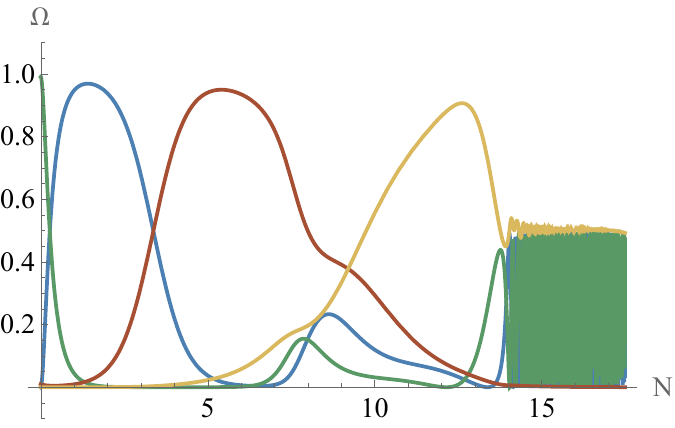}
    \includegraphics[width=0.6\textwidth]{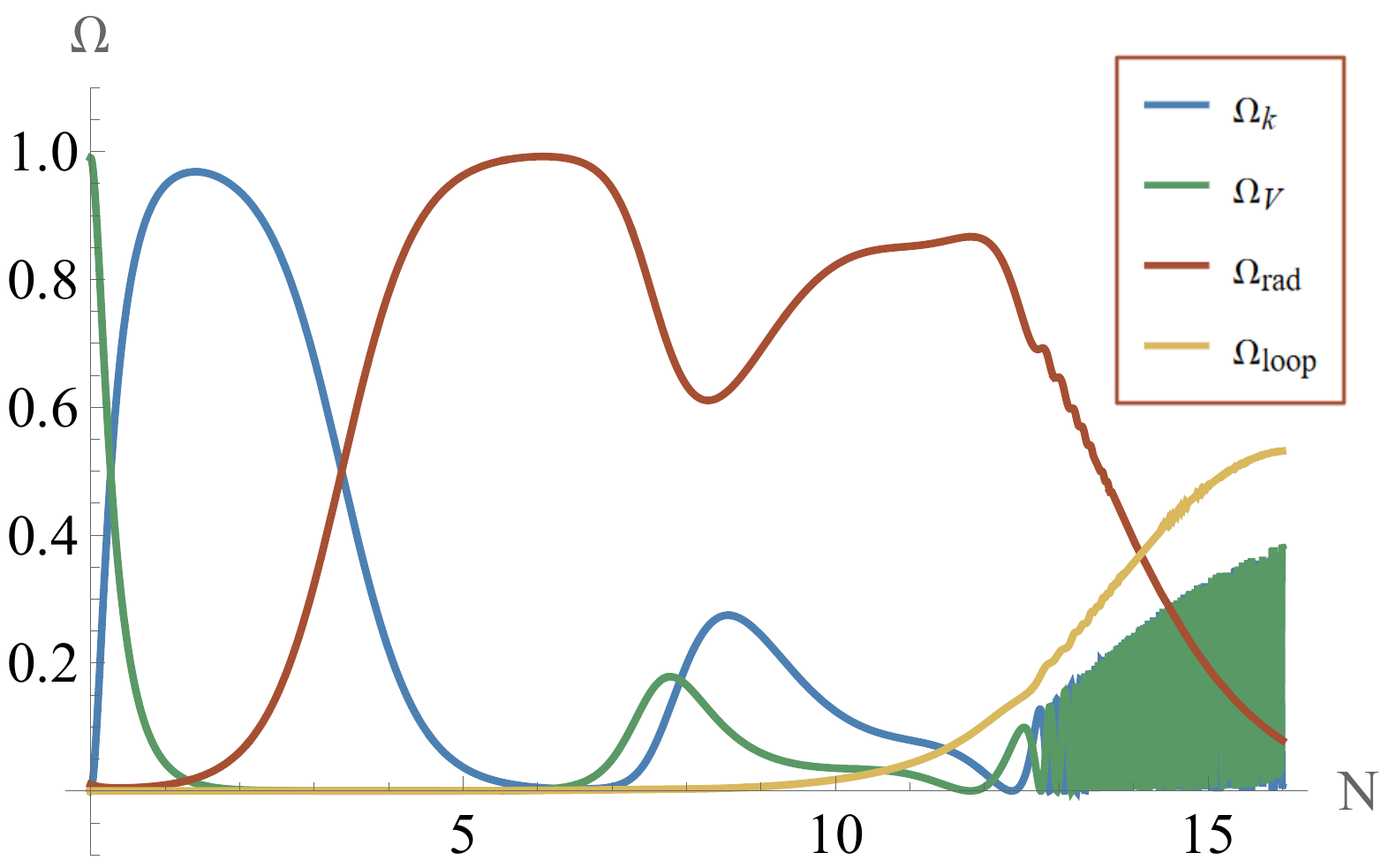}
    \caption{Evolution of the fractional energy densities of the dynamical system with NS5-strings ($\beta = 1/6$) and radiation, with $\Phi_0 = 6 M_p$, $X^2_0 = 0$, $W_0^2=10^{-2}$ and $Y^2_0=1-X_0^2-Z_0^2-W_0^2$. The initial conditions for the loop fraction differ in the 3 plots. \textit{Top:}  $Z_0^2 = 1 \times 10^{-3}$.  \textit{Middle}: $Z_0^2 = 1 \times 10^{-5}$.
\textit{Bottom}: $Z_0^2 = 1 \times 10^{-7}$.}
    \label{fig:NS5 rad}
\end{figure}

The case of NS5-strings is interesting, as the field would not overshoot even in the absence of radiation. As can be seen in Fig. \ref{fig:NS5 rad}, in this case there are two different behaviours depending on the initial hierarchy between radiation and loops. If $W_0^2/Z_0^2 \lesssim \O(10^2)$, the system behaves effectively as in the absence of radiation, with the radiation energy density barely getting to dominate in the early evolution and then decaying quickly. On the other hand, if the hierarchy is much higher, say $W_0^2/Z_0^2 \gtrsim \O(10^4)$, there is a pronounced growth of the fraction of energy density in loops after the field starts oscillating. The intermediate regime $\O(10^2) \lesssim W_0^2/Z_0^2 \lesssim \O(10^4)$ still goes back to the case without radiation at late times, but the system spends a few efolds in an epoch of radiation domination at early times.

\subsection{Dissipation due to GW emission}\label{sec:GW dissipation}

The present analysis has neglected the loss in energy density of loops due to gravitational wave emission, which is a continuous process. An argument for which one should be careful in doing this is the following, adapted from \cite{SanchezGonzalez:2025uco}. 
The typical power of emission of gravitational wave radiation is given by:
\begin{equation}
P_\GW \simeq \frac{\Gamma}{8\pi} \left(\frac{\mu}{M_p}\right)^2\,,
\end{equation}
where $\Gamma$ is a numerical factor depending on the loop configuration, which is generically fixed at $\Gamma \simeq 50$. A loop of length $\ell$ (and mass $\mu\times \ell$) will have a typical decay time in gravitational waves:
\begin{equation}
\tau_\GW \simeq 8\pi \frac{\ell M_p^2}{\Gamma  \mu}\,.
\end{equation}
While for F-strings, depending on initial conditions, the argument of \cite{SanchezGonzalez:2025uco} shows that the loss is essentially negligible in a Hubble time, for effective strings things are different.  
In particular, for NS5-strings, the tension goes as $\mu \sim M_s^2 \V^{2/3}\sim M_p^2/\V^{1/3}$, so that:
\begin{equation}
\tau_\GW \sim 8\pi \frac{ \V^{1/3}}{\Gamma} \, \ell\,.
\end{equation}
At early times, GW emission is expected to be irrelevant, and so in one Hubble time, $t_H=1/H$, the evolution of the length $\ell$ of the string is controlled just by the change in its tension $\mu$ as $\ell \sim \mu^{-1/2}\sim \V^{1/6}/M_p$ 
so that:
\begin{equation}
\label{eq:tau/t}
\frac{\tau_\GW}{t_H} \simeq 8\pi \frac{\sqrt{\V}} {\Gamma}\, \frac{H}{M_p}\,.
\end{equation}
Depending on the initial values of $H$ and $\V$, this fraction may or may not be relevant as the field starts evolving. What is surely interesting is that, unlike the case of F-strings, the emission process becomes more and more relevant as the volume approaches its minimum. A simple way to see this, is tracking the behaviour of the ratio \eqref{eq:tau/t} along the evolution of the system. As the volume rolls down its potential, it quickly enters a kinating epoch. During such a period, $\V \sim a^3 \sim e^{3N}$, while, on the other hand, $H \sim a^{-3}\sim e^{-3N}$.
Therefore, if the kinating epoch lasts $N_{\rm k}$ efolds:
\begin{equation}
\frac{\tau_\GW}{t_H} = \left(\frac{\tau_\GW}{t_H}\right)_{\rm in} e^{-\frac{3}{2} N_{\rm k}}\,.
\end{equation}
After that, the system enters the loop tracking regime. One can use the second Friedman    equation \eqref{eq:H dot} with $\rho_{\rm rad} = 0$ to retrieve how $\V$ and $H$ evolve with the scale factor in that period. For NS5-strings ($\beta = 1/6$), one gets $\V \sim a^{1/2} \sim e^{N/2}$, while $H \sim a^{-\frac{37}{24}} \sim e^{-\frac{37}{24}N}$. 
Supposing the transition between kination and loop tracking is instantaneous, and the loop-tracker period lasts $N_{\rm loop}$ efolds, one finds:
\begin{equation}
\frac{\tau_\GW}{t_H} = \left(\frac{\tau_\GW}{t_H}\right)_{\rm in} e^{-\frac{3}{2} N_{\rm k}-\frac{31}{24}N_{\rm loop}}\,,
\end{equation}
implying that the loss of energy into GWs becomes increasingly relevant at late times. 
This is to be contrasted with \cite{SanchezGonzalez:2025uco} where the focus on F-strings implied that GW production was more efficient at early times. 

By neglecting energy loss into GWs, we are assuming that the parameters in the above equation, in particular the initial string length and compactification volume, are such that $\tau_\GW\gg t_H$ at all times. If that is not the case, the dependence of the loop size on the GW emission has to be included. As argued in \cite{Ghoshal:2025tlk}, the size of the loop decreases due to GW emission. This effect may be more or less efficient than the growth of the string length due to the decrease in tension. For NS5-strings, the effect can be much larger at late times, implying that the physical string length could decrease in time, further reducing $\tau_\GW$.
An efficient emission of GWs would actually help in solving the overshoot problem, as the emitted radiation contributes in slowing the field down as it rolls towards its minimum. On the other hand, the fraction of residual energy density in loops at the minimum would be surely affected. Therefore, to assess the late-time evolution of the system, a more detailed analysis including the energy loss should be carried out. Such a study should take into account the dependence of the size of the loops on the produced radiation, which influences both the frequency of the emitted GWs and their amplitude. A detailed analysis of the GW spectrum for a network of cosmic superstrings has recently been conducted in \cite{Ghoshal:2025tlk}, in a similar stringy embedding as the present paper. We leave the precise analysis of the efficiency of the production and the GW spectrum of the setup presented here for future work.

\section{Oscillating tension}\label{sec:oscillating}

In this section we study the evolution of an isolated circular string as the modulus oscillates around the minimum of its potential, with the aim of understanding the final stage of the evolution outlined above. For simplicity we take the minimum to be quadratic:
\begin{equation}
\label{eq:quadratic minimum}
    V(\phi)=\frac{1}{2}m^2 \phi^2\,,
\end{equation}
and the field to follow:
\begin{equation}
\ddot{\phi}+3 H \dot{\phi}=-V_\phi\,.
\label{eq:KG}
\end{equation}
In this section we will neglect the backreaction of the individual string loop on the spacetime.

The dynamics of the system presents two distinct time scales: the scalar oscillation time, $T_\phi\equiv m^{-1}$, and the string oscillation time, $T_R \equiv \epsilon$. In the limit when the string oscillates may times during a single $\phi$ oscillation:
\begin{equation}
T_\phi \gg T_R\quad\Leftrightarrow\quad \epsilon m\ll 1 \,,
\label{eq:WKB}
\end{equation}
one can approximate $a^2 \dot{R}^2$ by its average $\langle a^2 \dot{R}^2\rangle =1/2$, integrate \eqref{eq:epsilonEOM} and reduce the dynamics to: 
\begin{equation}
    \ddot{R}+ \left( 2 H+\frac{\dot{\mu}}{2 \mu}\right )\dot{R}\,+ \frac{R}{\epsilon_0^2} \frac{\mu}{\mu_0}=0\,.
\label{eq:REOMeff}
\end{equation}
In this regime one may define the variable:
\begin{equation}
\sigma = a\ R\ \mu^\frac{1}{4}\,,
\label{eq:sigma}
\end{equation}
that obeys the following equation of motion:
\begin{equation}
    \ddot{\sigma}+\left(\frac{\mu}{\mu_0 \epsilon_0^2}+\frac{3}{16}\frac{ \dot{\mu}^2}{\mu^2}-\frac{1}{4}\frac{\ddot{\mu}}{\mu}-\frac{\ddot{a}}{a}-\frac{1}{2}\frac{\dot{a} \dot{\mu}}{a \mu}\right)\sigma=0\,.
\label{eq:Hill}
\end{equation}
Noting that when the modulus oscillates around its minimum, so does the string tension, \eqref{eq:Hill} is a Hill equation. The oscillating tension may open up a rich dynamics for the circular strings, in particular the possibility of resonances.

As is often the analysis of Mathieu/Hill's equations, it is convenient to rescale the time variable such that the period of the field oscillation is exactly $\pi$. We therefore define $\tau \equiv \frac{m}{2} t$ such that \eqref{eq:Hill} becomes:
\begin{equation}
\sigma''+\left(4 \frac{\mu/\mu_0}{m^2 \epsilon_0^2}+\frac{3}{16}\frac{ \mu'^2}{\mu^2}-\frac{1}{4}\frac{\mu''}{\mu}-\frac{a''}{a}-\frac{1}{2}\frac{a' \mu'}{a \mu}\right)\sigma=0\,.
\label{eq:Hilltau}
\end{equation}

Equation \eqref{eq:Hilltau} is an accurate description of the circular string dynamics, governed by \eqref{eq:epsilonEOM} and \eqref{eq:REOM}, in the regime $m \epsilon_0 \ll 1$. In this limit one may approximate \eqref{eq:Hilltau} by keeping only the first term inside the brackets, allowing for an approximate WKB solution of the form:
\begin{equation}
\sigma(\tau)= \sigma_0 \left(\frac{\mu_0}{\mu(\tau)}\right)^{1/4} \cos\left[ \frac{2}{m \epsilon_0}\int_0^\tau d \tilde{\tau} \left(\frac{\mu_0}{\mu(\tilde{\tau})} \right)^{1/2}+\delta\right],
\label{eq:solWKB}
\end{equation}
where $\delta$ is a phase fixed by the initial conditions. We see that the amplitude of the oscillations is modulated by the time variation of the tension. We stress that \eqref{eq:solWKB}  holds both in flat space and in the presence of gravity. Note also that this result is independent of whether the variation of the string tension is monotonic (as in \cite{Conlon:2024uob}) or oscillating, provided it  varies sufficiently slowly, cf. \eqref{eq:WKB}. In the WKB regime the physical radius of the string scales as $R_{\rm max}\propto a^{-1} \mu^{-1/2}$, and so it grows if:
\begin{equation}
    \frac{d}{d t}(a \mu^{1/2})<0 \Leftrightarrow \frac{\dot{\mu}}{\mu}<-2 H\,,
\end{equation}
that is, if the growth condition of \cite{Conlon:2024uob} is respected.

\subsection{Flat spacetime}

\begin{figure}
\centering
\includegraphics[width=0.7\textwidth]{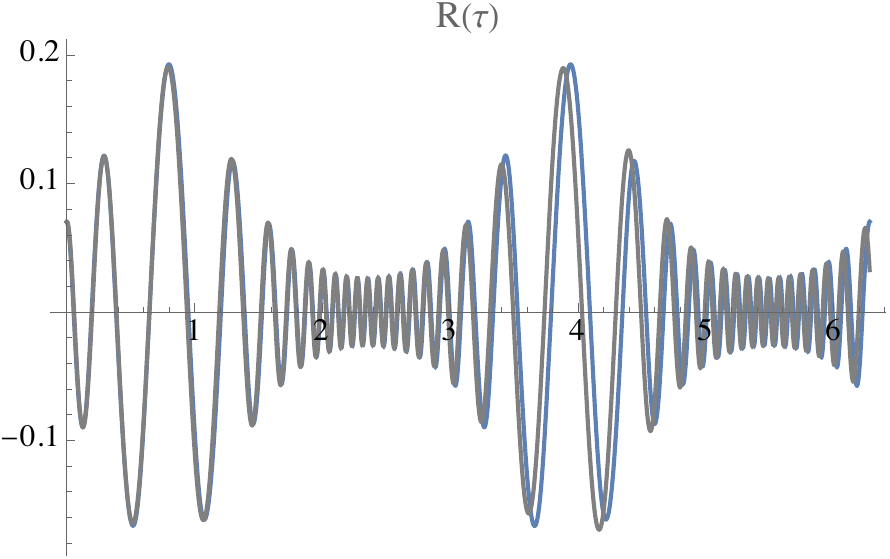}
\caption{Exact solution (blue) and WKB approximation (grey) for a loop with a time varying tension in flat space with $m=1$, $\epsilon_0=0.07$ and  $\beta \phi_0=\sqrt{2/3}$. The parameters have been chosen such that the difference between the WKB and the exact solution can be observed.}
\label{fig:WKBflat}
\end{figure}

In order to build some physical intuition for the behaviour of the strings with an oscillating tension, let us first analyse the flat space solutions of \eqref{eq:epsilonEOM} and \eqref{eq:REOM}. In flat space $a={\rm const}$, $H=0$ and the scalar motion is harmonic as there is no friction:
\begin{equation}
\phi=\phi_0 \cos(m t)=\phi_0 \cos( 2 \tau)\,.
\end{equation}
The tension then evolves as:
\begin{equation}
\mu=\mu_0\ e^{-\sqrt{6}\beta\,   \phi_0 \cos(2 \tau)}\,.
\end{equation}
From \eqref{eq:sigma} and \eqref{eq:solWKB} one then sees that the radius of the string follows:
\begin{equation}
R(\tau)\propto \mu(\tau)^{-1/2}\,,
\end{equation}
showing clearly that, in this regime, the string radius is modulated by the scalar field oscillations but crucially that no resonance is present. A numerical example of this solution is presented in Fig. \ref{fig:WKBflat}.

\begin{figure}
\centering
\includegraphics[width=0.7\textwidth]{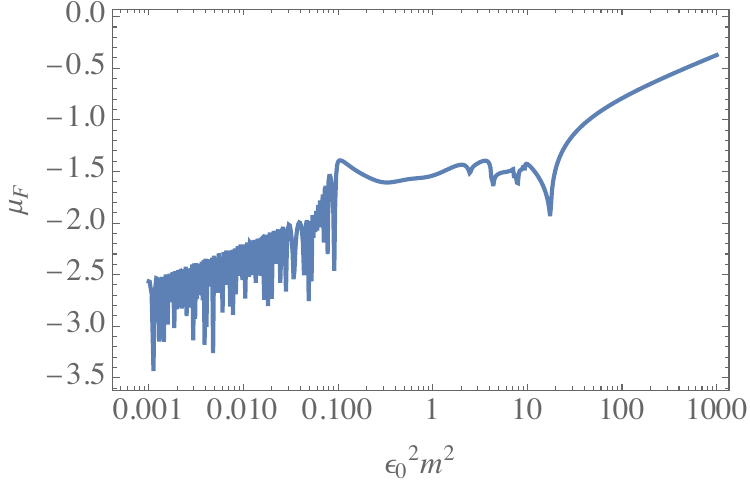}
\caption{No instability in flat space: growth exponent as a function of the timescale of the problem $\epsilon_0^2 m^2$ (obtained from a numerical analysis of \eqref{eq:epsilonEOM}-\eqref{eq:REOM} with $\beta \phi_0=\sqrt{2/3}$.)}
\label{fig:noInst}
\end{figure}

\begin{figure}
    \centering
    \includegraphics[width=0.49\textwidth]{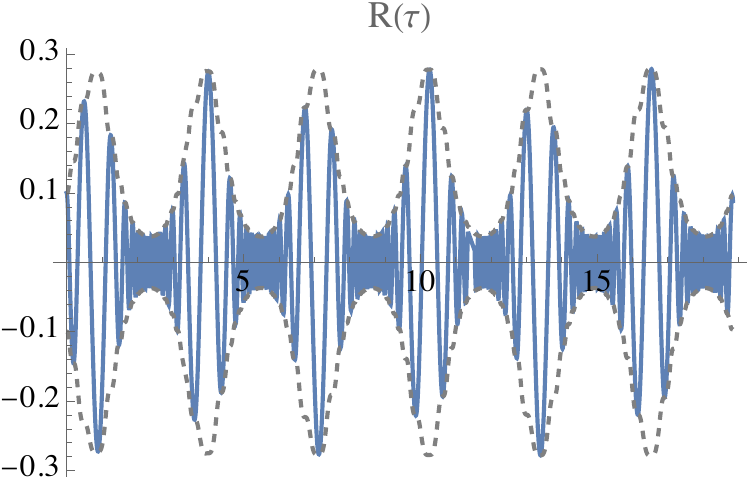}
    \includegraphics[width=0.49\textwidth]{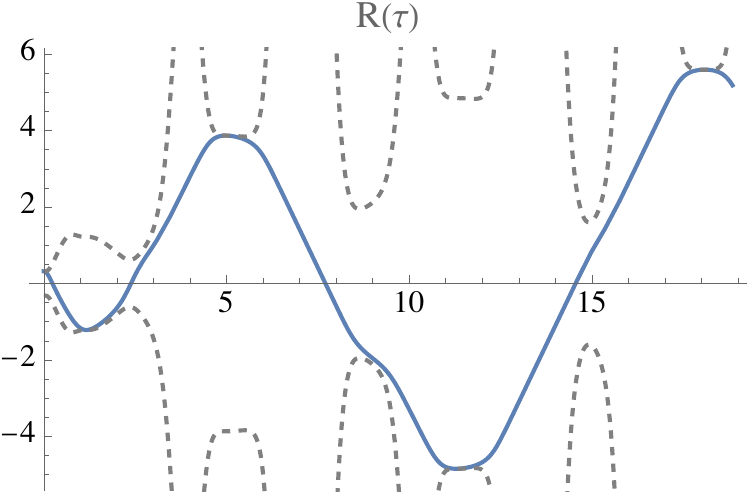}
    \includegraphics[width=0.49\textwidth]{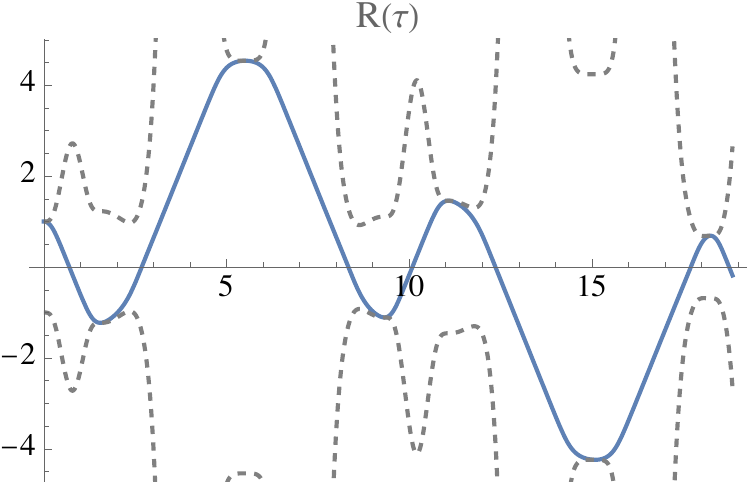}
    \includegraphics[width=0.49\textwidth]{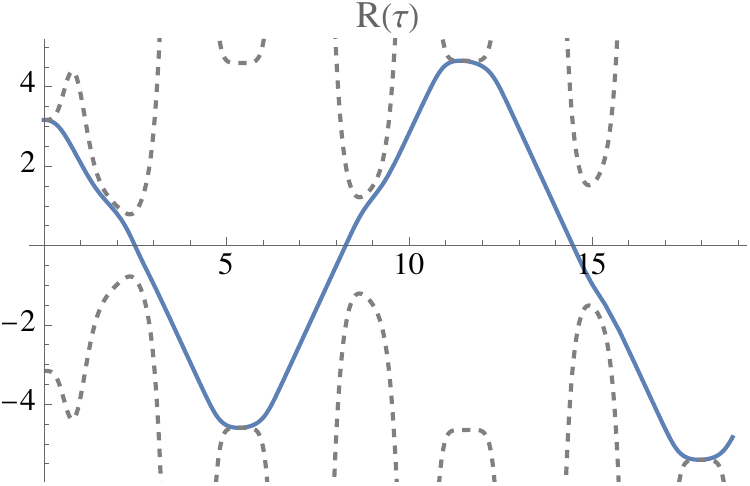}
    \includegraphics[width=0.49\textwidth]{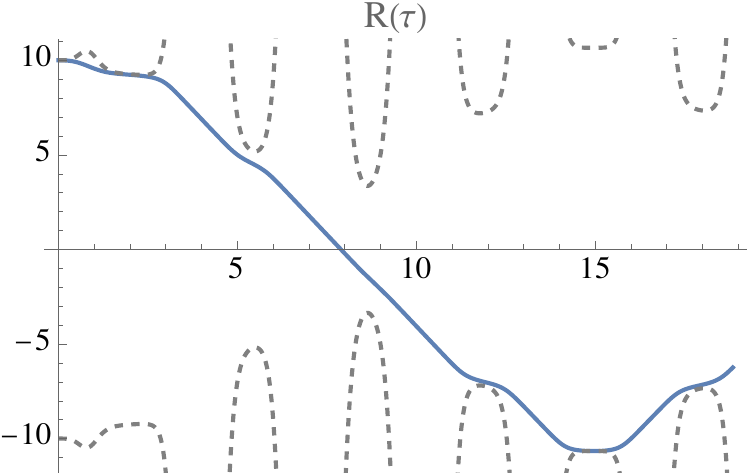}
\caption{No sign of instability for $\beta \phi_0=\sqrt{2/3}$. Top to bottom: $\epsilon_0^2 m^2=\{10^{-2},10^{-1},1,10, 10^2\}$. The dashed lines represent $\epsilon(\tau)$.}
\label{fig:examplesFlatSpace}
\end{figure}

Having seen that in the WKB regime the system presents no instability, we adapt the standard methods of Floquet's analysis (see e.g. \cite{Amin:2014eta}) to the current problem and look for instabilities when \eqref{eq:WKB} is not satisfied. Defining the Floquet exponent $\mu_F\in \mathbb{R}$ via:
\begin{equation}
R(\tau)= e^{\mu_F\,\tau} f(\tau)\,,
\end{equation}
where $f(\tau)$ is a periodic function, one may numerically determine it for any given choice of $(\beta \phi_0, \epsilon_0, m)$. In Fig. \ref{fig:noInst} we present one such scan, where it is evident that $\mu_F<0$, implying that there is no exponential growth of the closed string loop.\footnote{The exponent is calculated after a complete oscillation of the scalar field. Unlike in the linear Mathieu/Hill equation, where the value of the exponent is independent of the time when it is computed, for this non-linear two variable system we find some weak dependence.} We have repeated the analysis for different values of the input parameters and found repeatedly that $\mu_F<0$. Several representative trajectories for $ \beta \phi_0 =\sqrt{2/3}$ are shown in Fig. \ref{fig:examplesFlatSpace}, none of which exhibits exponential behaviour.

\subsection{Curved spacetime}

The addition of gravity brings with it Hubble friction that dampens the scalar field oscillation and consequently decreases the time variation of the string loop tension.

Provided the strings do not significantly contribute to the overall energy density, the Friedmann equation reads:
\begin{equation}
H^2=\frac{1}{3 M_p^2}\left(\frac{1}{2}\dot{\phi}^2 +V(\phi)\right).
\label{eq:Friedmann}
\end{equation}
As above, we are interested in a regime where the scalar is oscillating around the minimum of its potential. If the potential is well approximated by a quadratic term, the solutions to \eqref{eq:KG} and \eqref{eq:Friedmann} are of the form:
\begin{equation}
\begin{split}
\phi(\tau)&\approx\frac{\phi_0}{\tau} \cos( 2\tau)\,,\\
H(\tau)&\approx \frac{H_0}{\tau}\,,
\end{split}
\end{equation}
implying the following time variation of the string tension:
\begin{equation}
\mu=\mu_0 \ e^{- \sqrt{6}\beta\,\frac{\phi_0}{\tau} \cos( \tau)}\,.
\end{equation}
We see that in the late time limit, the variation of the tension becomes less significant. Taking into account this time dependence, one can rewrite the solution \eqref{eq:solWKB} in terms of the string radius as:
\begin{equation}
R(\tau)\propto \frac{\mu(\tau)^{-1/2}}{a(\tau)}\,.
\end{equation}
Thus the variation of the string maximal radius caused by the oscillating tension is dampened by the expansion of spacetime. Once again we found no resonances but only some mild enhancement/suppression of the string radius depending on the string initial conditions.

This absence of resonant behaviour for an isolated string with oscillating tension does not preclude interesting cosmological behaviour for a diluted gas of strings, as we have seen in the previous sections.

\section{Discussion and conclusions}
\label{sec:conclusion}

In this work we have analysed the dynamics of loops of cosmic superstrings in the 4D EFT of type IIB string theory as the volume modulus controlling their tension evolves towards the minimum of its potential. 

The first part of this work has been devoted to studying whether the rolling volume mode actually settles in its minimum or overshoots it causing a decompactification limit. In fact, the typical LVS scalar potential for the canonically normalised volume mode $\Phi$ \eqref{eq:LVS potential} features a minimum and a maximum. If $\Phi$ reaches the minimum with too much kinetic energy, the field will overshoot the maximum and run away to infinity. In the absence of any background fluid, this overshoot problem is present for generic initial conditions. The canonical way to overcome this problem is to introduce a fraction of radiation in the system. Depending on the specific embedding of the model, this solution may or may not be well justified.
Therefore, following the work of \cite{SanchezGonzalez:2025uco}, we studied how the inclusion of a string loop fluid in the dynamical system affects the overshooting problem. 

We first analysed the behaviour of the  dynamical system including cosmic string loops in the absence of a radiation background. We performed a stability analysis of the fixed points, summarised in the stability map in Fig. \ref{fig:stab_no_fluid}. For an exponential potential with $\lambda = 3$, as for the approximated LVS potential \eqref{eq:Lvs exponential}, we found that the final attractor of a system with only F-strings ($\beta = 1/2$) is the loop tracker $\mathcal L$, confirming the result of \cite{SanchezGonzalez:2025uco}. For $\beta = 1/2$, the kinetic energy of the field is too high to stop at the minimum, and the system overshoots. 
On the other hand, for effective strings coming from D3-branes wrapping an internal 2-cycle ($\beta = 1/3$) or NS5-branes wrapping a 4-cycle ($\beta = 1/6$), the final attractor of the system would be the mixed tracker $\mathcal T_1$. As the convergence towards this fixed point is slow, the system still spends a long time hovering around $\mathcal L$, even if this critical point is unstable. Hence, whether the field overshoots or converges to the minimum is still dependent on the fraction of kinetic energy at $\mathcal L$. For $\beta = 1/3$, the situation is qualitatively similar to the F-string case, with a kinetic energy that is too high at the minimum to avoid overshooting. 
For $\beta= 1/6$, the NS5-string loops are so effective at absorbing energy from the field that they can indeed prevent it from overshooting. 
In fact, for NS5-strings, at $\mathcal L$ the kinetic energy of the field constitutes only a small fraction of the total energy density, $\Omega_{\rm k} = \beta^2 \simeq 0.03$, while the fraction of energy density in NS5-string loops gets over $97 \%$, as shown in Fig. \ref{fig:NS5 no rad}. 
Moreover, when the field starts oscillating around the minimum, the tension of the strings stops decreasing. Therefore, the loops start redshifting as matter, just like the field. This means that the ratio of their energy densities stays constant throughout the period of early matter domination, provided that the strings do not lose energy through GW or particle emission. The most interesting aspect is that this fraction is always rather high, ranging from $25\%$ to $60\%$. This is very promising, as the signal in GW emission is proportional to the energy density in string loops. The asymptotic fraction of energy density in NS5-string loops shows an interesting dependence on the initial conditions of the field $\Phi_0$ and on the initial loop energy density fraction $\Omega^{(0)}_{\rm loop}$, as shown in Fig. \ref{fig:final density}. 

It is worth noting that in the presence of multiple non-interacting string fluids, like F-strings, D3-strings and NS5-strings, with similar initial energy densities, the dominant contribution is still expected to arise from NS5-strings, as these redshift the slowest.

We then turned on a radiation background fluid in the system and repeated our analysis. Again, we performed a stability study of the fixed points and the results are shown in the map in Fig. \ref{fig:stability rad}. In a similar fashion to what was observed in the absence of radiation, the F-string attractor for $\lambda =3$ is the loop tracker \cite{SanchezGonzalez:2025uco}. 
Nevertheless, depending on the initial fraction of energy density in radiation and in loops, the system spends quite a long time around the radiation tracker $\mathcal S$. This slows the field down and, in the absence of string loops, would prevent it from overshooting. However, if the loops are present, the system will eventually move towards $\mathcal L$, again leading to overshooting, unless the field reaches its minimum while the system is transitioning from $\mathcal S$ to $\mathcal L$. In this case, the loop energy density may still grow  as shown in Fig. \ref{fig:F string rad}. 
Moreover, as argued above, as the field oscillates around the minimum, $\Omega_{\rm loops}/\Omega_\Phi$ stays constant. However, radiation redshifts faster than matter, and so the radiation density drops while $\Omega_\Phi$ and $\Omega_{\rm loop}$ grow. This leads to a growth in the energy density of cosmic string loops \textit{after} $\Phi$ settles into its minimum, until it reaches its asymptotic value and stabilises there.

A similar effect can be observed for effective strings as well. For D3-string ($\beta = 1/3$), we have essentially the same behaviour as for F-strings, while the case of NS5-strings ($\beta = 1/6$) is more nuanced. Depending on the ratio of the initial energy densities of radiation and loops, $\Omega_{\rm rad}^{(0)}/\Omega_{\rm loop}^{(0)}$, we can either go back to the case with no radiation ($\Omega_{\rm rad}^{(0)}/\Omega_{\rm loop}^{(0)}\lesssim 10^4$) or mimic the F-string case ($\Omega_{\rm rad}^{(0)}/\Omega_{\rm loop}^{(0)}\gtrsim 10^4$), as shown in Fig. \ref{fig:NS5 rad}.

The latter part of this work focused on studying an isolated string with an oscillating tension. We have shown that, when the field reaches the minimum, the oscillations in the string tension induce a modulation of its radius, though our analysis has not revealed any instabilities.

An interesting development would be the evaluation of the spectrum of GWs emitted by the string loops, both during the period of decreasing tension and during the oscillating one. In particular, given that NS5-strings can reach over $97\%$ of the total energy density of the universe, they might emit powerful (hence, observable) GW signals at high frequencies. A fundamental step that needs to be done towards the conclusive assessment of the spectrum is understanding the production mechanism for the string loops. This would give information about the initial distribution of the number densities and lengths of the various string species, essential prerequisite for the study of GW emission. Finally, an equally interesting future direction would be the calculation of the spectrum of emitted particles. In particular, effective strings are expected to have couplings to axions, closed- and open-string moduli and gauge bosons living on branes. A cross-correlation between the spectra of emitted GWs and particles would be of great phenomenological relevance.  

In conclusion, our results demonstrate that cosmic superstrings with a time varying tension may play an important role in the early universe with consequences that we are just starting to unveil.

\vspace{0.7cm}

\paragraph*{\textbf{Acknowledgments:}}

We thank Joseph P. Conlon, Noelia Sanchez Gonzalez and Gonzalo Villa for useful discussions. The work of LB, MC and FGP contributes to the COST Action COSMIC WISPers CA21106, supported by COST (European Cooperation in Science and Technology).

\appendix

\section{Stability analysis}

In this appendix we gather the analytic conditions for the linear stability of the various fixed points, summarized in Figs. \ref{fig:stab_no_fluid} and \ref{fig:stability rad}, following the standard procedure outlined in e.g. \cite{Copeland:1997et}.

\subsection{Loops without radiation}

\begin{itemize}
    \item $\mathcal{K}_+$: $\lambda>2$ and $\beta>1$
    \item $\mathcal{M}$:  $0 < \lambda < 2$ and $\beta > (-2 + \lambda^2)/\lambda$
    \item $\mathcal{L}$: $\lambda > \frac{1+\beta^2}{\beta}$ and $\beta<1$
    \item $\mathcal{T}_1$: $\frac{\beta}{2}+\sqrt{2+\frac{\beta^2}{4}}<\lambda<\frac{1+\beta^2}{\beta}$ and $\beta<1$
\end{itemize}
\subsection{Loops with radiation}

\begin{itemize}
    \item $\mathcal{K}_+$: unstable
    \item $\mathcal{M}$:  $0 < \lambda < 2$ and $\beta > \frac{-2 + \lambda^2}{\lambda}$
    \item $\mathcal{L}$: $\beta<\frac{1}{\sqrt{3}}$ and $\lambda> \frac{1+\beta^2}{\beta}$
    \item $\mathcal{T}_1$: $\left(\frac{\beta}{2}+\frac{\sqrt{8+\beta^2}}{2}<\lambda<\frac{1+\beta^2}{\beta}\right.$ and $\left. \beta<\frac{1}{\sqrt{6}}\right)$ or $ \left(4 \beta <\lambda<\frac{1+\beta^2}{\beta}\right.$ and $\left. \beta>\frac{1}{\sqrt{6}}\right)$
    \item $\mathcal{S}$: $\beta>\frac{1}{\sqrt{6}}$ and $2\sqrt{\frac{2}{3}}<\lambda< 4 \beta$
    \item $\mathcal{LR}$: always a stable valley
    \item $\mathcal{T}_2$: $\beta > \frac{1}{\sqrt{3}}$ and $\lambda>4\beta$
    \item $\mathcal{F}$:  unstable
\end{itemize}

\newpage
\bibliographystyle{JHEP}
\bibliography{references}

\end{document}